\documentstyle[psfig,epsf,axodraw]{article}
\begin{document}
\renewcommand{\arraystretch}{1.1}
\newcommand{\flin}[2]{\ArrowLine(#1)(#2)}
\newcommand{\glin}[3]{\Photon(#1)(#2){2}{#3}}
\newcommand{\nl}{\nonumber \\}
\newcommand{\lsim}
{\mathrel{\raisebox{-.3em}{$\stackrel{\displaystyle <}{\sim}$}}}
\newcommand{\sof}{\SetOffset}
\newcommand{\vj}[4]{{\sl #1~}{\bf #2 }
 \ifnum#3<100 (19#3) \else (#3) \fi #4}
\newcommand{\ej}[3]{{\bf #1~}
 \ifnum#2<100 (19#2) \else (#2) \fi #3}
\newcommand{\ap}[3]{\vj{Ann.~Phys.}{#1}{#2}{#3}}
\newcommand{\app}[3]{\vj{Acta~Phys.~Pol.}{#1}{#2}{#3}}
\newcommand{\cmp}[3]{\vj{Commun. Math. Phys.}{#1}{#2}{#3}}
\newcommand{\cnpp}[3]{\vj{Comments Nucl. Part. Phys.}{#1}{#2}{#3}}
\newcommand{\cpc}[3]{\vj{Comp. Phys. Commun.}{#1}{#2}{#3}}
\newcommand{\epj}[3]{\vj{Eur. Phys. J.}{#1}{#2}{#3}}
\newcommand{\hpa}[3]{\vj{Helv. Phys.~Acta}{#1}{#2}{#3}} 
\newcommand{\ijmp}[3]{\vj{Int. J. Mod. Phys.}{#1}{#2}{#3}}
\newcommand{\jetp}[3]{\vj{JETP}{#1}{#2}{#3}}
\newcommand{\jetpl}[3]{\vj{JETP Lett.}{#1}{#2}{#3}}
\newcommand{\jmp}[3]{\vj{J. Math. Phys.}{#1}{#2}{#3}}
\newcommand{\jp}[3]{\vj{J. Phys.}{#1}{#2}{#3}}
\newcommand{\lnc}[3]{\vj{Lett. Nuovo Cimento}{#1}{#2}{#3}}
\newcommand{\mpl}[3]{\vj{Mod. Phys. Lett.}{#1}{#2}{#3}}
\newcommand{\nc}[3]{\vj{Nuovo Cimento}{#1}{#2}{#3}}
\newcommand{\nim}[3]{\vj{Nucl. Instr. Meth.}{#1}{#2}{#3}}
\newcommand{\np}[3]{\vj{Nucl. Phys.}{#1}{#2}{#3}}
\newcommand{\pl}[3]{\vj{Phys. Lett.}{#1}{#2}{#3}}
\newcommand{\prp}[3]{\vj{Phys. Rep.}{#1}{#2}{#3}}
\newcommand{\pr}[3]{\vj{Phys. Rev.}{#1}{#2}{#3}}
\newcommand{\prl}[3]{\vj{Phys. Rev. Lett.}{#1}{#2}{#3}}
\newcommand{\ptp}[3]{\vj{Prog. Theor. Phys.}{#1}{#2}{#3}}
\newcommand{\rpp}[3]{\vj{Rep. Prog. Phys.}{#1}{#2}{#3}}
\newcommand{\rmp}[3]{\vj{Rev. Mod. Phys.}{#1}{#2}{#3}}
\newcommand{\rnc}[3]{\vj{Rivista del Nuovo Cim.}{#1}{#2}{#3}}
\newcommand{\sjnp}[3]{\vj{Sov. J. Nucl. Phys.}{#1}{#2}{#3}}
\newcommand{\sptp}[3]{\vj{Suppl. Prog. Theor. Phys.}{#1}{#2}{#3}}
\newcommand{\zp}[3]{\vj{Z. Phys.}{#1}{#2}{#3}}
\newcommand{\jop}[3]{\vj{Journal of Physics} {\bf #1} (#2) #3}
\newcommand{\ibid}[3]{\vj{ibid.} {\bf #1} (#2) #3}
\newcommand{\hep}[1]{{\sl hep--ph/}{#1}}
\begin{center}
{\Large \bf Topics on four-fermion Physics at electron-positron colliders
\footnote{Invited talk presented at QFTHEP'2000, Tver, Russia.}} \\
\vspace{4mm}
Roberto Pittau\\
Dipartimento di Fisica Teorica, 
Universit\`a di Torino, Italy\\
INFN, Sezione di Torino, Italy
\end{center}
\begin{abstract}
\noindent I review the most recent progresses in the calculation 
of four-fermion processes in $e^+ e^-$ collisions.
\end{abstract}
\section*{Introduction}

Final state four-fermion processes represent an important 
ingredient when studying high energy $e^+$ $e^-$ collisions.
They enter in the analysis of the $Z$-peak observables at LEP,
as (mainly) QED pair corrections to 2-fermion processes. 
Such a set of contributions, together with the full set of 
electroweak (EW) loop  corrections, allows very stringent 
tests the Standard Model (SM) at the radiative level.

At LEP2 energies, all relevant signatures such as $WW$, $ZZ$, single-$W$,
single-$Z$  and Higgs production manifest themselves 
as four-fermion final states.
These same processes are also relevant for precision measurements
at the Linear Collider (LC) and as a SM background to searches.

In the following, based on the CERN Report of ref. \cite{wshop}, 
I review 
the most recent achievements in the computation of four-fermion processes.
Table \ref{tab0} contains a list of contributing codes.
Emphasis is put here on the available tools for estimating 
the theoretical errors to be associated with the four-fermion
observables, in view of the final LEP analysis, but also on the 
improvements needed at the LC.

The first two sections are devoted to $W$-pair and single-$W$ signatures, 
while, in the last two parts, I cover topics on four-fermion production 
plus 1 visible photon and $Z$-pair final states.
\begin{table}[thb]
\begin{center}
\begin{tabular}{|l|l|}
\hline
Code & ~~~~Authors \\
\hline
{\tt BBC} \cite{bbc} &
$~~~$ F. Berends, W Beenakker and A. Chapovsky
\\
{\tt CompHEP} \cite{cop}&
$~~~$ E. Boos, M. Dubinin and V. Ilyin
\\
 {\tt GENTLE} \cite{gee}&
$~~~$  D. Bardin, A. Olchevsky and T. Riemann
\\
 {\tt GRACE} \cite{gre}&
$~~~$  Y. Kurihara, M. Kuroda and Y. Shimizu
\\
{\tt KORALW/YFSWW} \cite{kow}&
$~~~$  S. Jadach, W. Placzek, M. Skrypek, B. Ward and Z. Was
\\
{\tt NEXTCALIBUR} \cite{ner}&
$~~~$  F. Berends, C. G. Papadopoulos and R. Pittau
\\
{\tt PHEGAS/HELAC} \cite{phs}&
$~~~$  C. G. Papadopoulos
\\
{\tt RACOONWW} \cite{raw} &
$~~~$  A. Denner, S. Dittmaier, M. Roth and D. Wackeroth
\\
{\tt SWAP} \cite{swp}&
$~~~$ G. Montagna, M. Moretti, O. Nicrosini, A. Pallavicini
 and F. Piccinini
\\
{\tt WPHACT} \cite{wpt}&
$~~~$ E. Accomando, A. Ballestrero and E. Maina
\\
{\tt WRAP} \cite{swp} &
$~~~$ G. Montagna, M. Moretti, O. Nicrosini, M. Osmo
  and F. Piccinini
\\
{\tt WTO} \cite{wto} & 
$~~~$  G. Passarino
\\
{\tt YFSZZ} \cite{yfz}& 
$~~~$ S. Jadach, W. Placzek and B.F.L. Ward
\\
{\tt ZZTO} \cite{zzo} & 
$~~~$ G. Passarino \\ \hline
\end{tabular}
\caption{\label{tab0} Contributing programs.}
\end{center}
\end{table}
\section{$W$-pair production}
When collecting data at $\sqrt{s}=$ 189 GeV, the LEP2 collaborations
observed a deficit in the number of events, with respect 
to the SM predictions. 
This fact triggered a re-analysis of the
available tools for calculating the total cross section $\sigma_{WW}$.
At that time, a theoretical 2\% error band was assigned to this observable, 
two times bigger than the experimental error.
The estimate of the theoretical error was based on the 
{\tt {GENTLE/4FAN}} inclusion of QED Initial State Radiation (ISR), 
without any attempt to take EW 
contributions into account. Therefore, it was immediately clear 
that the computation of the genuine EW effects was needed to 
match the experimental accuracy. 
On the other hand a full four-fermion one-loop EW calculation 
was (and still is) beyond reach, and including only the $WW$-like 
diagrams violates gauge invariance.
 The solution to this problem is represented by the so called Double 
Pole Approximation (DPA) \cite{dpa}.
The DPA isolates the poles at the complex squared 
masses, with gauge invariant residues which are then projected 
onto the respective on-shell gauge invariant counterparts.
The projection is from the off-shell phase space to the 
on-shell phase space. Even though such a procedure is strictly gauge invariant,
the projection procedure is not unique. However, 
the ambiguity is small, namely 
${\cal O}(\frac{\alpha}{\pi}\frac{\Gamma_W}{M_W})$.

For example, in the case of a single unstable particle, 
the fully re-summed amplitude can be rewritten as follows
\begin{eqnarray}
   {\cal M}^\infty 
     &=& \frac{W(p^2,\omega)}{p^2-\tilde{M}^2}\,\sum_{n=0}^{\infty}
         \,\Biggl( \frac{-\tilde{\Sigma}(p^2)}{p^2-\tilde{M}^2} 
           \Biggr)^n 
      =\ \frac{W(p^2,\omega)}{p^2-\tilde{M}^2+\tilde{\Sigma}(p^2)}
         \nonumber \\[1mm]
     &=& \frac{W(M^2,\omega)}{p^2-M^2}\,\frac{1}{Z(M^2)} + \Biggl[  
         \frac{W(p^2,\omega)}{p^2-\tilde{M}^2+\tilde{\Sigma}(p^2)}
         - \frac{W(M^2,\omega)}{p^2-M^2}\,\frac{1}{Z(M^2)} \Biggr] 
    \nonumber \\ \nonumber \\
  &&M^2-\tilde{M}^2+\tilde{\Sigma}(M^2) = 0, \quad\quad
  Z(M^2) = 1+\tilde{\Sigma}'(M^2), 
\end{eqnarray}
where $\tilde{M}$ and $M$ are the bare mass and the complex pole
of the instable particle, and $Z$ the wave-function factor.
The first term is the gauge invariant single-pole residue
(on-shell production and decay of the unstable particle). 
The second term has no pole and can be in principle expanded in powers of $\,p^2-M^2$.

\begin{figure}[thb]
\begin{center}
  \unitlength .7pt\small\SetScale{0.7}
  \begin{picture}(200,110)(0,-10)
    \Text(-30,92)[lc]{(a)}
    \ArrowLine(43,58)(25,40)        \Text(7,40)[lc]{$e^+$}
    \ArrowLine(25,90)(43,72)        \Text(7,92)[lc]{$e^-$}
    \Photon(50,65)(150,95){1}{12}   \Text(100,105)[]{$W$}  
    \Photon(50,65)(150,35){1}{12}   \Text(100,25)[]{$W$}
    \ArrowLine(155,98)(175,110)     \Text(190,110)[rc]{$f_1'$}
    \ArrowLine(175,80)(155,92)      \Text(190,80)[rc]{$\bar{f}_1$}
    \ArrowLine(175,50)(155,38)      \Text(190,50)[rc]{$\bar{f}_2'$}
    \ArrowLine(155,32)(175,20)      \Text(190,20)[rc]{$f_2$}
    \DashLine(75,115)(75,15){5}     \Text(40,5)[]{production}
    \DashLine(125,115)(125,15){5}   \Text(150,5)[]{decays}
    \GCirc(50,65){10}{1}
    \GCirc(100,80){10}{0.8}
    \GCirc(100,50){10}{0.8}
    \GCirc(150,95){10}{1}
    \GCirc(150,35){10}{1}
  \end{picture}
\end{center}
\begin{center}
  \unitlength .7pt\small\SetScale{0.7}
  \begin{picture}(120,35)(0,0)
  \Text(-30,92)[lc]{(b)}
  \ArrowLine(30,50)( 5, 95)
  \ArrowLine( 5, 5)(30, 50)
  \Photon(30,50)(90,80){2}{6}
  \Photon(30,50)(90,20){2}{6}
  \GCirc(30,50){10}{0}
  \Vertex(90,80){1.2}
  \Vertex(90,20){1.2}
  \ArrowLine(90,80)(120, 95)
  \ArrowLine(120,65)(105,72.5)
  \ArrowLine(105,72.5)(90,80)
  \Vertex(105,72.5){1.2}
  \ArrowLine(120, 5)( 90,20)
  \ArrowLine( 90,20)(105,27.5)
  \ArrowLine(105,27.5)(120,35)
  \Vertex(105,27.5){1.2}
  \Photon(105,27.5)(105,72.5){2}{4.5}
  \put(92,47){$\gamma$}
  \put(55,73){$W$}
  \put(55,16){$W$}
  \end{picture}
  \quad\quad
  \begin{picture}(120,100)(0,0)
  \ArrowLine(30,50)( 5, 95)
  \ArrowLine( 5, 5)(30, 50)
  \Photon(30,50)(90,80){2}{6}
  \Photon(30,50)(90,20){2}{6}
  \Vertex(60,35){1.2}
  \GCirc(30,50){10}{0}
  \Vertex(90,80){1.2}
  \Vertex(90,20){1.2}
  \ArrowLine(90,80)(120, 95)
  \ArrowLine(120,65)(105,72.5)
  \ArrowLine(105,72.5)(90,80)
  \Vertex(105,72.5){1.2}
  \ArrowLine(120, 5)(90,20)
  \ArrowLine(90,20)(120,35)
  \Photon(60,35)(105,72.5){2}{5}
  \put(87,46){$\gamma$}
  \put(63,11){$W$}
  \put(38,22){$W$}
  \put(55,73){$W$}
  \end{picture}
  \quad\quad 
  \begin{picture}(160,100)(0,0)
  \ArrowLine(30,50)( 5, 95)
  \ArrowLine( 5, 5)(30, 50)
  \Photon(30,50)(90,80){-2}{6}
  \Photon(30,50)(90,20){2}{6}
  \Vertex(60,65){1.2}
  \GCirc(30,50){10}{0}
  \Vertex(90,80){1.2}
  \Vertex(90,20){1.2}
  \ArrowLine(90,80)(120, 95)
  \ArrowLine(120,65)(105,72.5)
  \ArrowLine(105,72.5)(90,80)
  \Vertex(105,27.5){1.2}
  \ArrowLine(120, 5)(90,20)
  \ArrowLine(105,27.5)(120,35)
  \ArrowLine(90,20)(105,27.5)
  \Photon(60,65)(105,27.5){-2}{5}
  \put(84,55){$\gamma$}
  \put(63,78){$W$}
  \put(40,68){$W$}
  \put(55,16){$W$}
  \end{picture}
  \\[2ex]
  \unitlength .7pt\small\SetScale{0.7}
  \begin{picture}(240,100)(0,0)
  \ArrowLine(30,50)( 5, 95)
  \ArrowLine( 5, 5)(30, 50)
  \Photon(30,50)(90,80){2}{6}
  \Photon(30,50)(90,20){2}{6}
  \GCirc(30,50){10}{0}
  \Vertex(90,80){1.2}
  \Vertex(90,20){1.2}
  \ArrowLine(90,80)(120, 95)
  \ArrowLine(120,65)(105,72.5)
  \ArrowLine(105,72.5)(90,80)
  \ArrowLine(120, 5)( 90,20)
  \ArrowLine( 90,20)(120,35)
  \Vertex(105,72.5){1.2}
  \PhotonArc(120,65)(15,150,270){2}{3}
  \put(55,73){$W$}
  \put(55,16){$W$}
  \put(99,47){$\gamma$}
  \DashLine(120,0)(120,100){6}
  \PhotonArc(120,35)(15,-30,90){2}{3}
  \Vertex(135,27.5){1.2}
  \ArrowLine(150,80)(120,95)
  \ArrowLine(120,65)(150,80)
  \ArrowLine(120, 5)(150,20)
  \ArrowLine(150,20)(135,27.5)
  \ArrowLine(135,27.5)(120,35)
  \Vertex(150,80){1.2}
  \Vertex(150,20){1.2}
  \Photon(210,50)(150,80){2}{6}
  \Photon(210,50)(150,20){2}{6}
  \ArrowLine(210,50)(235,95)
  \ArrowLine(235, 5)(210,50)
  \GCirc(210,50){10}{0}
  \put(177,73){$W$}
  \put(177,16){$W$}
  \end{picture}
  \end{center}
\caption{\label{fig2} Factorizable (a) and non factorizable (b)
contributions to $W$-pair production.}
\end{figure}
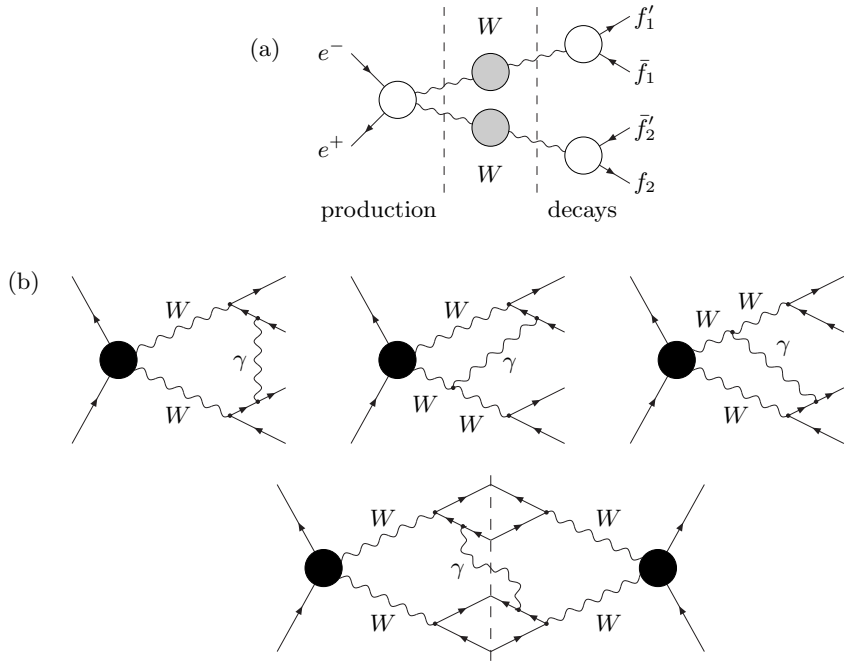

 Applying DPA to $W$-pair production means that only the double-pole 
residues of the two resonances are considered, 
and one-loop EW contributions included there,
for which only (available) on-shell corrections are needed.
The corrections to be included fall in two different classes, namely
factorizable contributions, in which the production, propagation and decay
steps are clearly separated, and non-factorizable contributions, in which
a photon with energy $E_\gamma \lsim \Gamma_{W}$ is emitted
(see figure \ref{fig2}).
The DPA is not reliable at the $W$-pair threshold, where  
the background diagrams get important.
The expected DPA uncertainty above threshold
is of the order ${\cal O}\left(\frac{\alpha}{\pi}\frac{\Gamma_W}{M_W} 
{\rm ln( \cdots)}\right)\,< 0.5\%$, in fact, when
$\sqrt{s} > 2\,M_W + n \Gamma_W$ with $n= {\cal O}(3-5)$,  
the background diagrams are of the order
$\sim \frac{\alpha}{\pi}\frac{\Gamma_W}{\sqrt{s}-M_W}\, 
{\rm ln( \cdots)} \sim 0.1\%$.

Very far away from resonance, the DPA cannot be used any more.

\begin{figure}[thb]
\centerline{\psfig{figure=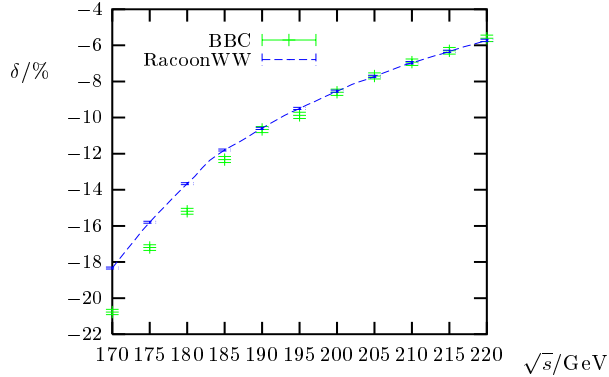,height=5.0cm,width=8cm}}
\caption{\label{fig3} 
Relative ${\cal O}({\alpha})$ corrections to 
$\sigma(e^+ e^- \to \nu_\mu\mu^+\tau^-\bar \nu_\tau)$.}
\end{figure}

\begin{figure}[thb]
\vskip -1.6cm
\centerline{\psfig{figure=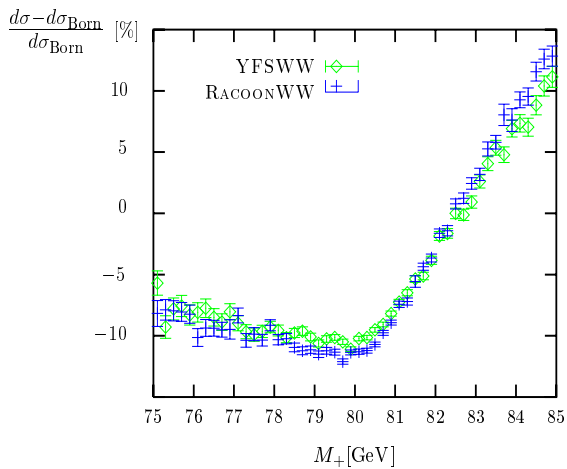,height=20.0cm,width=13cm}}
\vskip -12cm
\caption{\label{fig5} 
Distribution in the $W^+$ invariant mass at $\sqrt{s}=200\,{\rm GeV}$
for the $u\bar d\mu^-\bar \nu_\mu$ final state.}
\end{figure}

At LEP2 energies, the inclusion of the DPA formalism in {\tt RACOONWW},  
{\tt BBC} and {\tt YFSWW} allows to lower the theoretical uncertainty 
on $\sigma_{WW}$ from 2\% to 0.5 \%, in much better agreement with the data.
In figures \ref{fig3} and \ref{fig5} and table \ref{fig4} we show examples 
of comparisons among the codes.  

\vskip 0.3cm
\begin{table}[thb]
\begin{center}
\begin{tabular}{|c|c|c|c|}
\hline
\multicolumn{2}{|c|}{\bf no cuts}&
\multicolumn{2}{|c|}{\bf$\sigma_{\mathrm{tot}}[\mathrm{fb}]$}\\
\hline
final state & program & Born & best \nl
\hline\hline
& {\tt YFSWW3} & 219.770(23) & 199.995(62) \nl
$\nu_\mu\mu^+\tau^-\bar\nu_\tau$
& {\tt RacoonWW} & 219.836(40) & 199.551(46) \nl
\cline{2-4}
& (Y--R)/Y & $-0.03(2)$\% &  0.22(4)\% \nl
\hline\hline
& {\tt YFSWW3} & 659.64(07) & 622.71(19) \nl
$ u\bar d\mu^-\bar\nu_\mu$
& {\tt RacoonWW} & 659.51(12) & 621.06(14) \nl
\cline{2-4}
& (Y--R)/Y & $0.02(2)$\% &  0.27(4)\% \nl
\hline\hline
& {\tt YFSWW3} & 1978.18(21) & 1937.40(61) \nl
$ u\bar d s\bar c$
& {\tt RacoonWW} & 1978.53(36) & 1932.20(44) \nl
\cline{2-4}
& (Y--R)/Y & $-0.02(2)$\% &  0.27(4)\% \nl
\hline
\end{tabular}
\end{center}
\caption{\label{fig4} Total cross section
at $\sqrt{s}=200\,{\rm GeV}$ without cuts.}
\end{table}
In conclusion, with the help of the DPA, a theoretical 
accuracy at the level of 0.5 \% on $\sigma_{WW}$ is reached, as 
required by the LEP2 collaborations \cite{wshop}.
The error decreases with increasing energy, giving the following 
estimates of the theoretical uncertainty
on $\sigma_{\rm WW}$
\begin{itemize}
\item[ ] 0.4 \% at $\sqrt{s}= 200$ GeV,~
         0.5 \% at $\sqrt{s}= 180$ GeV,~
         0.7 \% at $\sqrt{s}= 170$ GeV.
\end{itemize}
A theoretical uncertainty of the order of 1 \% must be assigned to 
the distributions.

A full four-fermion one-loop EW calculation is still missing, 
but it is required for high precision measurements at the LC.
\section{Single-$W$ production}
\begin{figure}[thb]
  \begin{picture}(110,70)(0,0)
  \SetOffset(50,0)
  \ArrowLine(50,50)(0,100)
  \ArrowLine(100,100)(50,50)
  \ArrowLine(0,0)(50,50)
  \ArrowLine(50,50)(100,0)
  \ArrowLine(100,70)(50,50)
  \ArrowLine(50,50)(100,30)
  \GCirc(50,50){15}{0.5}
  \Text(10,100)[lc]{$e^+$}
  \Text(10,0)[lc]{$e^-$}
  \Text(110,100)[lc]{$\bar \nu_e$}
  \Text(110,70)[lc]{$\bar f_2$}
  \Text(110,30)[lc]{$f_1$}
  \Text(110,0)[lc]{$e^-$}
  \Text(65,-20)[t]{$s$-channel}
\end{picture}
\begin{picture}(110,70)(0,0)
  \SetOffset(150,0)
  \ArrowLine(50,50)(0,100)
  \ArrowLine(100,100)(50,50)
  \ArrowLine(100,70)(50,50)
  \ArrowLine(50,50)(100,30)
  \Photon(50,50)(50,10){2}{7}
  \ArrowLine(0,0)(50,10)
  \ArrowLine(50,10)(100,0)
  \GCirc(50,50){15}{0.5}
  \Text(10,100)[lc]{$e^+$}
  \Text(10,-5)[lc]{$e^-$}
  \Text(110,100)[lc]{$\bar \nu_e$}
  \Text(110,70)[lc]{$\bar f_2$}
  \Text(110,30)[lc]{$f_1$}
  \Text(110,-5)[lc]{$e^-$}
  \Text(65,-20)[t]{$t$-channel}
  \Text(60,20)[lc]{$t$}
  \Text(-40,50)[lc]{$+$}
  \end{picture}
\vskip  1.cm
\caption{\label{fig6} $s$-channel and $t$-channel diagrams in 
single-$W$ production.}
\end{figure}
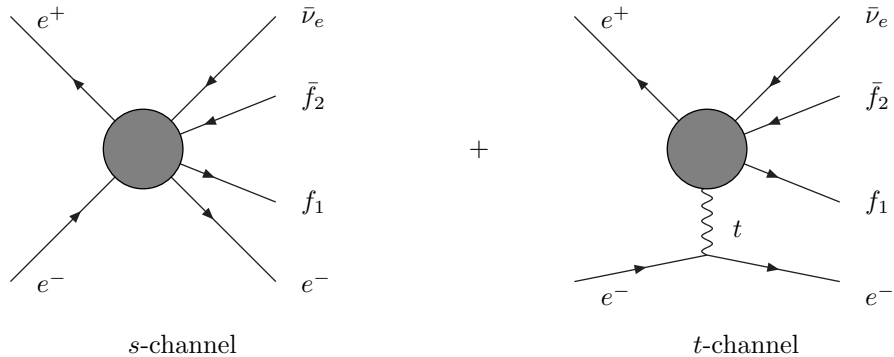

The experimental single-$W$ signature is better defined with 
the help of table \ref{tab1}.
\begin{table}[thb]
\begin{center}
\begin{tabular}{|c|c|c|}
\hline
Process & diagrams & cuts \\
\hline
$ee\nu\nu$ & $t$-channel only & 
$E(e^+) > 20\,GeV, |\cos\theta(e^+)| < 0.95$\\  
$e\nu\mu\nu$ & $t$-channel only  & $E_(\mu^+)  > 20\,GeV$     \\
$e\nu\tau\nu$ & $t$-channel only & $E_(\tau^+) > 20\,GeV$     \\
$e\nu u d$ & $t$-channel only    & $M(ud) > 45 GeV$           \\
$e\nu c s$ & $t$-channel only    & $M(cs) > 45 GeV$           \\
\hline
\end{tabular}
\caption{\label{tab1} Possible single-$W$ processes.
$|\cos\theta(e^-)| > 0.95$.}
\end{center}
\end{table}
The contributing Feynman diagrams can be split into $s$-channel 
and $t$-channel amplitudes, as depicted in figure \ref{fig6}. 
\vskip  1.5cm
Only the $t$-channel contribution is included 
in the definition of the single-$W$ processes. This set 
is explicitly drawn, for the  $e\nu u d$ 
final state, in figure \ref{fig7}.
The reason of this diagram based definition is that it allows an 
easy combination of different processes from different experiments.
Notice that, in order to preserve gauge invariance, {\em all} 
$t$-channel diagrams must be included, not only those where a 
$W$ boson is produced.

Since in the limit of vanishing electron masses the $t$-channel 
diagrams blow up in the forward region, the computation of single-$W$ 
processes is a challenge from a technical point of view. 
In addition, including by hand a width for the $W$ boson 
breaks the $U(1)$ gauge invariance, so that gauge preserving 
schemes must be applied, such as the Fixed Width (FW) 
and the Imaginary Fermion Loop (IFL) approaches \cite{fwandfl}.

Generally speaking, the technical questions are well under control,
both for total cross sections and distributions, as shown by the 
comparisons in tables \ref{tab2}-\ref{tab3} and figure \ref{fig8}.

As for a realistic modelling of this process, the main difficulties
are due to the presence of different scales. 
Usually, when studying high energy processes, part of the higher order 
corrections can be reabsorbed in the Born approximation
by using the so-called $G_F$ scheme. 
In such a scheme $G_F$, $M_Z$ and $M_W$ are 
input parameters, while the weak mixing angle and 
$\alpha_{EM}$ are derived quantities:
\begin{eqnarray}
s^2_W ~=~ 1 - {M^2_W}/{M^2_Z}\,,~~~~ 
\alpha_{EM} ~=~  \sqrt{2}\,\,\frac{G_F\,M_W^2\,s^2_W}{\pi} \,.
\end{eqnarray}
In the presence of low $t$-channel scales such an approach fails,
since the choice $\alpha(t \sim 0) \sim 1/137$ is certainly
more appropriate for the $t$-channel diagrams of figure \ref{fig6}.
 The question is therefore how to consistently include the running
of $\alpha_{EM}$ in single-$W$ four-fermion processes, without breaking
gauge invariance.

An exact and field-theoretically consistent solution is represented 
by the Exact Fermion-Loop (EFL) approach of ref. \cite{passa1}, 
where the whole, gauge invariant set 
of fermion one-loop corrections is taken into account, 
including the vertices.

In table \ref{tab4} the results of FW and EFL are compared for
different cuts on the outgoing electron.

There are also approximate solutions, such as  
{IFL}$_\alpha$ \cite{iflalpha} -- where
$\alpha(t)$ is used for the $t$-component and $\alpha_{G_F}$ 
for the $s$-component -- or the Modified Fermion Loop
(MFL) approach by {\tt NEXTCALIBUR}, that only 
includes the leading self-energy like corrections from the EFL 
plus an effective vertex to preserve the $U(1)$ gauge invariance.

IFL$_\alpha$ and MFL coincide numerically at LEP2 energies, but
slightly disagree with respect to {\tt EFL} (up to 2\% at LEP2).
Comparisons at different energies are presented 
in tables \ref{tab5} and \ref{tab6}.
\begin{figure}
\begin{picture}(300,328)(0,0)
  \SetScale{0.7}
  \SetWidth{1}
  \sof(80,230)
  \ArrowLine(50,120)(0,140)
  \ArrowLine(100,140)(50,120)
  \ArrowLine(0,0)(50,20)
  \ArrowLine(50,20)(100,0)
  \ArrowLine(100,110)(80,70)
  \ArrowLine(80,70)(100,30)
  \Photon(50,20)(50,70){2}{7}
  \Photon(50,70)(50,120){2}{7}
  \Photon(50,70)(80,70){2}{7}
  \Text(-14,98)[lc]{$e^+$}
  \Text(77,98)[lc]{$\bar \nu_e$}
  \Text(-14,0)[lc]{$e^-$}
  \Text(77,0)[lc]{$e^-$}
  \Text(77,77)[lc]{$\bar d$}
  \Text(77,21)[lc]{$u$}
  \Text(18,60)[lc]{$W$}
  \Text(11,26)[lc]{$\gamma, Z$}
  \sof(280,230)
  \ArrowLine(50,120)(0,140)
  \ArrowLine(65,126)(50,120)
  \ArrowLine(100,140)(65,126)
  \ArrowLine(0,0)(50,20)
  \ArrowLine(50,20)(100,0)
  \ArrowLine(100,110)(80,70)
  \ArrowLine(80,70)(100,30)
  \Photon(50,20)(50,120){2}{7}
  \Photon(65,126)(80,70){2}{7}
  \Text(77,77)[lc]{$\bar d$}
  \Text(77,21)[lc]{$u$}
  \Text(54,76)[lc]{$W$}
  \Text(11,66)[lc]{$\gamma,Z$}
  \Text(-14,98)[lc]{$e^+$}
  \Text(77,98)[lc]{$\bar \nu_e$}
  \Text(-14,0)[lc]{$e^-$}
  \Text(77,0)[lc]{$e^-$}
  \sof(80,115)
  \ArrowLine(50,120)(0,140)
  \ArrowLine(100,140)(50,120)
  \ArrowLine(0,0)(50,20)
  \ArrowLine(50,20)(100,0)
  \ArrowLine(100,110)(80,70)
  \ArrowLine(80,70)(100,30)
  \Photon(50,20)(50,120){2}{7}
  \Photon(18,132)(80,70){2}{7}
  \Text(77,77)[lc]{$\bar d$}
  \Text(77,21)[lc]{$u$}
  \Text(48,71)[lc]{$W$}
  \Text(21,56)[lc]{$Z$}
  \Text(-14,98)[lc]{$e^+$}
  \Text(77,98)[lc]{$\bar \nu_e$}
  \Text(-14,0)[lc]{$e^-$}
  \Text(77,0)[lc]{$e^-$}
  \sof(280,115)
  \ArrowLine(50,120)(0,140)
  \ArrowLine(100,140)(50,120)
  \ArrowLine(0,0)(50,20)
  \ArrowLine(50,20)(100,0)
  \ArrowLine(100,110)(80,70)
  \ArrowLine(80,70)(100,30)
  \Photon(50,20)(50,120){2}{7}
  \Photon(64,15)(80,70){2}{7}
  \Text(77,77)[lc]{$\bar d$}
  \Text(77,21)[lc]{$u$}
  \Text(40,41)[lc]{$W$}
  \Text(19,56)[lc]{$W$}
  \Text(-14,98)[lc]{$e^+$}
  \Text(77,98)[lc]{$\bar \nu_e$}
  \Text(-14,0)[lc]{$e^-$}
  \Text(77,0)[lc]{$e^-$}
  \sof(80,0)
  \ArrowLine(50,120)(0,140)
  \ArrowLine(100,140)(50,120)
  \ArrowLine(0,0)(50,20)
  \ArrowLine(50,20)(100,0)
  \ArrowLine(100,90)(50,90)
  \ArrowLine(50,90)(50,50)
  \ArrowLine(50,50)(100,50)
  \Photon(50,20)(50,50){2}{7}
  \Photon(50,90)(50,120){2}{7}
  \Text(77,72)[lc]{$\bar d$}
  \Text(77,26)[lc]{$u$}
  \Text(22,52)[lc]{$u$}
  \Text(10,20)[lc]{$\gamma,Z$}
  \Text(17,77)[lc]{$W$}
  \Text(-14,98)[lc]{$e^+$}
  \Text(77,98)[lc]{$\bar \nu_e$}
  \Text(-14,0)[lc]{$e^-$}
  \Text(77,0)[lc]{$e^-$}
  \sof(280,0)
  \ArrowLine(50,120)(0,140)
  \ArrowLine(100,140)(50,120)
  \ArrowLine(0,0)(50,20)
  \ArrowLine(50,20)(100,0)
  \ArrowLine(100,90)(50,50)
  \Line(50,90)(65,78)
  \ArrowLine(90,58)(100,50)
  \ArrowLine(50,50)(50,90)
  \Photon(50,20)(50,50){2}{7}
  \Photon(50,90)(50,120){2}{7}
  \Text(77,72)[lc]{$\bar d$}
  \Text(77,21)[lc]{$u$}
  \Text(22,52)[lc]{$d$}
  \Text(10,20)[lc]{$\gamma,Z$}
  \Text(17,77)[lc]{$W$}
  \Text(-14,98)[lc]{$e^+$}
  \Text(77,98)[lc]{$\bar \nu_e$}
  \Text(-14,0)[lc]{$e^-$}
  \Text(77,0)[lc]{$e^-$}
\end{picture}
\vskip 0cm
\caption{\label{fig7} $t$-channel diagrams for the $e\nu u d$
single-$W$ final state.}
\end{figure}
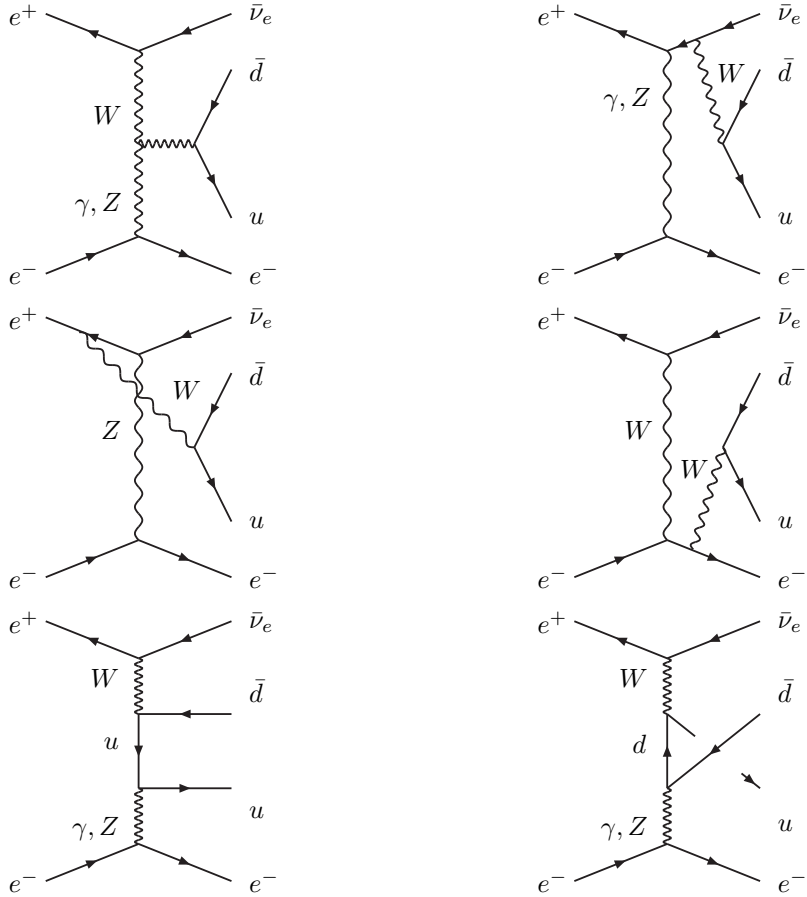
\begin{table}[thb]
\begin{center}
\begin{tabular}{|c|c|c|c|}
\hline
   & & &  \\
   & $\sqrt{s} = 183\,$GeV & $\sqrt{s} = 189\,$GeV & $\sqrt{s} = 200\,$GeV \\
   & & &  \\
\hline
   & & &  \\
{\tt NEXTCALIBUR}  & $26.483 \pm 0.041$ &  $29.679 \pm 0.047$
& $35.893 \pm 0.048$ \\
   & & &  \\
\hline
   & & &  \\
{\tt SWAP}      & $26.47 \pm 0.04$ &  $29.70  \pm 0.04$
& $35.93 \pm 0.05$ \\ 
   & & &  \\
\hline
\end{tabular}
\caption{\label{tab2} Cross-sections [fb] for $e^+e^- \to e^-\bar \nu_e \mu^+
\bar \nu_{\mu}$.}
\end{center}
\end{table}
\begin{table}[thb]
\begin{center}
\begin{tabular}{|c|c|c|c|}
\hline
   & & &  \\
   & $\sqrt{s} = 183\,$GeV & $\sqrt{s} = 189\,$GeV & $\sqrt{s} = 200\,$GeV \\
   & & &  \\
\hline
   & & &  \\
{\tt NEXTCALIBUR}  & $26.422 \pm 0.035$ &  $29.655 \pm 0.046$
& $35.954 \pm 0.052$ \\
   & & &  \\
\hline
   & & &  \\
{\tt SWAP}      & $26.3 \pm 0.2$ &  $29.6  \pm 0.2$
& $35.92 \pm 0.05$ \\ 
   & & &  \\
\hline
\end{tabular}
\caption{\label{tab3} Cross-sections [fb] for $e^+e^- \to e^-\bar \nu_e \tau^+
\bar \nu_{\tau}$.}
\end{center}
\end{table}

\begin{figure}[thb]
\vskip  0cm
\hskip -0.5cm
\psfig{figure=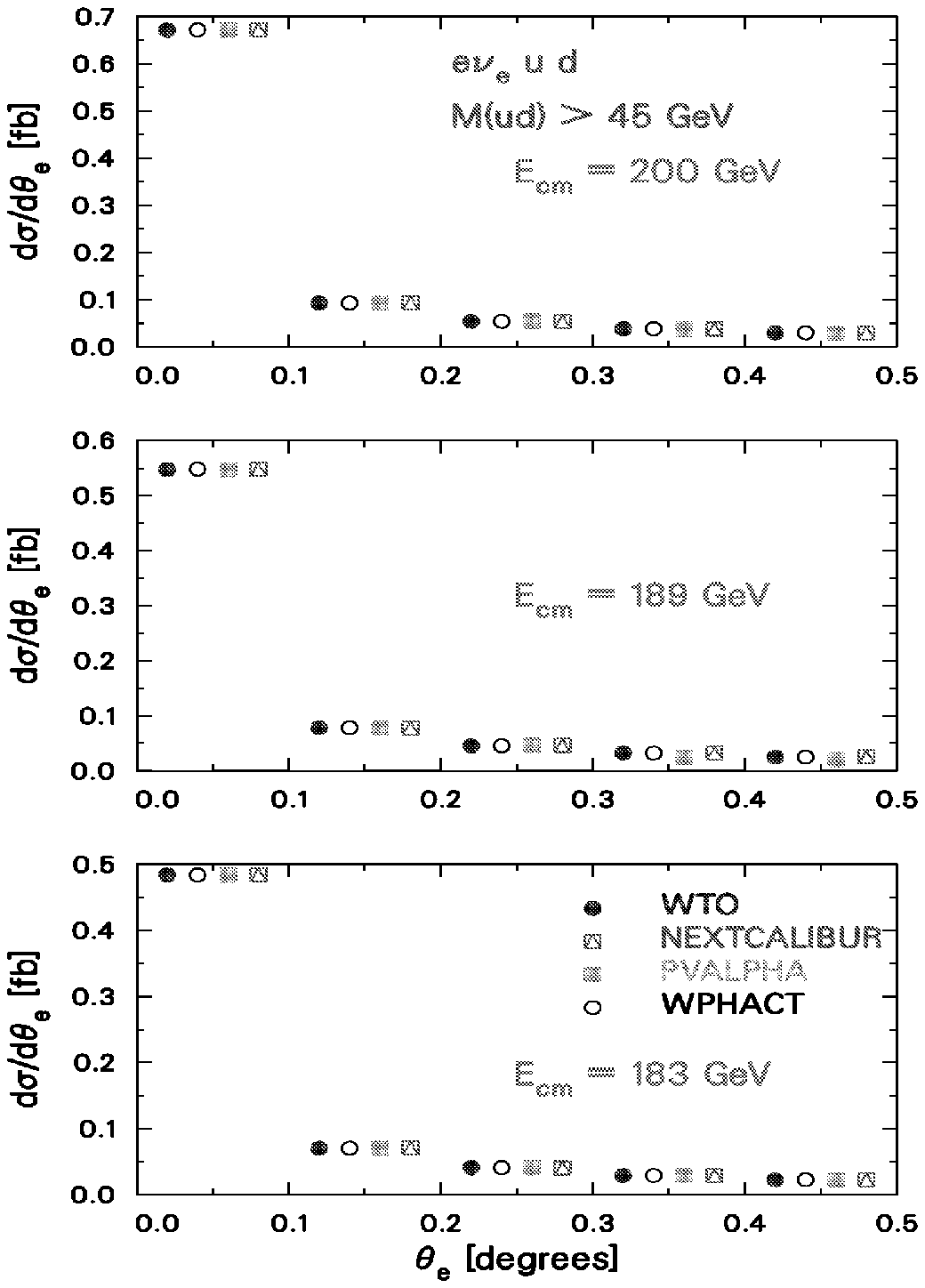,height=12.5cm,width=8cm}
\vskip -12.2cm
\hskip 7cm
\psfig{figure=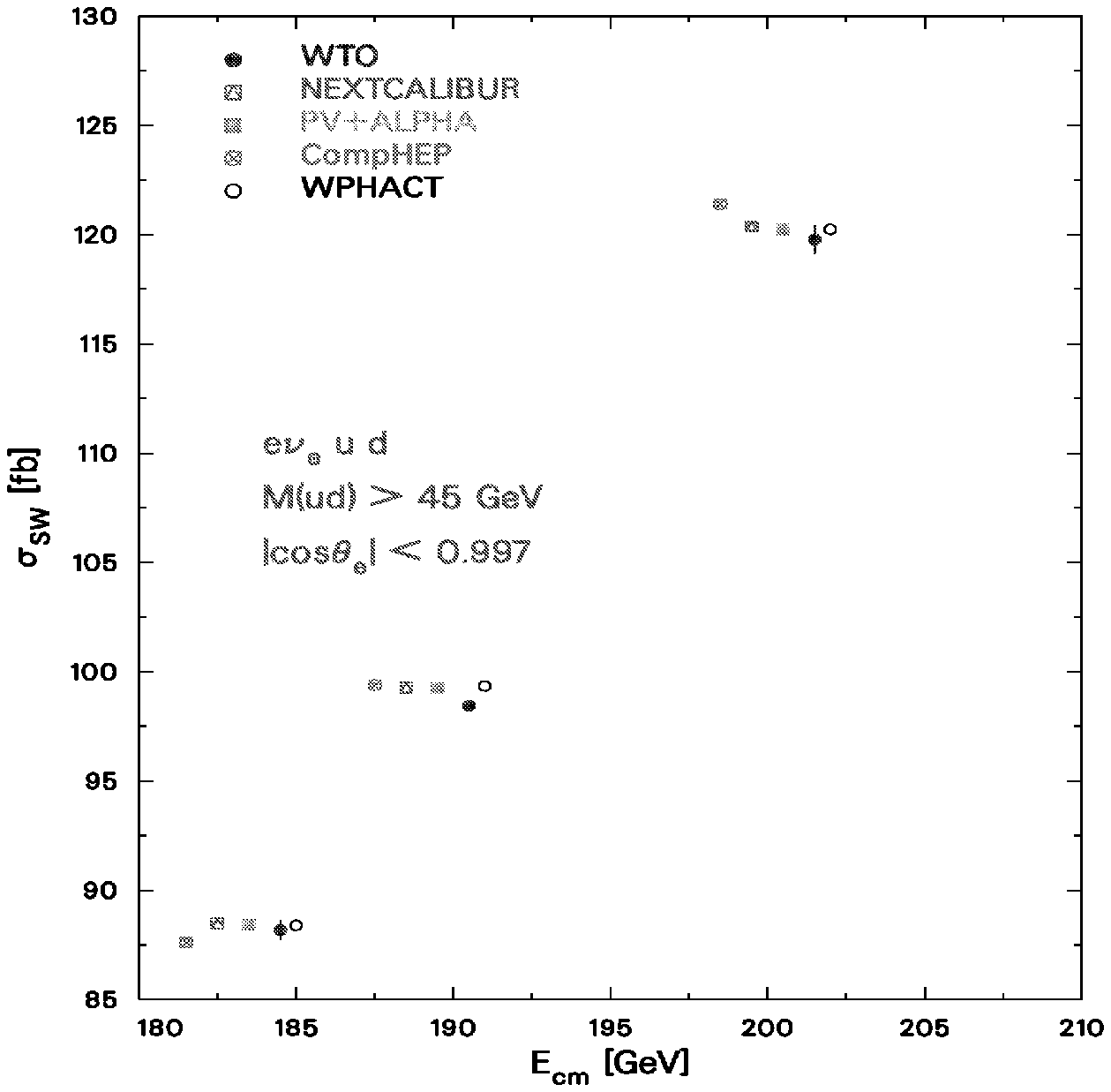,height=12.5cm,width=8cm}

\vskip -3.cm

\caption{\label{fig8} $\theta_e$ distribution and single-$W$ cross sections
for  $u \bar d e^- \bar \nu_e$.}
\end{figure}

\begin{table}[thb]
 \begin{center}
 \begin{tabular}{|c||c|c|c|}
 \hline
 $\theta_e\,$[Deg] &  FW  & EFL  & EFL/FW-1 (percent)  \\
 \hline
                                &           &           &               \\
   $0.0^\circ \div 0.1^\circ$   &  0.14154  & 0.13448   &  -4.99      \\
                                &           &           &               \\
 \hline
                                &           &           &               \\
   $0.1^\circ \div 0.2^\circ$   &  0.02113  & 0.02031   &  -3.88      \\
                                &           &           &               \\
 \hline
                                &           &           &               \\
   $0.2^\circ \div 0.3^\circ$   &  0.01238  & 0.01194   &  -3.55      \\
                                &           &           &               \\
 \hline
                                &           &           &               \\
   $0.3^\circ \div 0.4^\circ$   &  0.00880  & 0.00851   &  -3.30      \\
                                &           &           &               \\
 \hline
 \end{tabular}
\caption{\label{tab4} $d\sigma/d\theta_e$ [pb/degrees], 
from {\tt WTO}, for the process 
$e^+e^- \to e^- \bar \nu_e \nu_{\mu} \mu^+$, for $|\cos\theta_e| > 0.997$, 
$E_{\mu} > 15\,$GeV, and $|\cos\theta_{\mu}| < 0.95$. 
$\sqrt{s} = 183\,$ GeV.}
\end{center}
\end {table}

\begin{table}[thb]
 \begin{center}
 \begin{tabular}{|c||c|c|c|c|c|}
 \hline
            &      &     &    &    &                \\
 $\sqrt{s}$ &  FW  & IFL &IFL$_{\alpha}$ & EFL & EFL/FW-1  \\
            &      &     &    &    &   (percent)             \\
 \hline
            &      &     &          &         &                \\
 $183\,{\rm GeV}$ & 88.17(44) & 88.50(4)   & 83.26(5) & 83.28(6)  & -5.5(5)  \\
            &      &     &          &    &                \\
 $189\,{\rm GeV}$ & 98.45(25) & 99.26(4)   & 93.60(9) & 93.79(7)  & -4.7(3)  \\
            &      &     &      &    &                \\
 $200\,{\rm GeV}$ & 119.77(67)& 120.43(10) & 113.24(8)& 113.67(8) & -5.1(5)  \\
            &      &     &     &    &                \\
 \hline
 \end{tabular}
\caption{\label{tab5}
Cross section [fb] for the process $e^+e^- \to e^- \bar \nu_e u \bar d$.
$M(u\bar d) > 45\,$GeV, $|\cos\theta_e| > 0.997$. No ISR.}
 \end{center}
\end {table}

\begin{table}[thb]
\begin{center}
 \begin{tabular}{|c||c|c|c|c|c|}
 \hline
            &      &     &    &    &                \\
 $\sqrt{s}$ &  FW  & IFL &IFL$_{\alpha}$ & EFL & EFL/FW-1  \\
            &      &     &    &    &   (percent)             \\
 \hline
            &      &     &          &         &                \\
 $183\,{\rm GeV}$ &26.77(14)  &26.45(1)  & 24.90(1)  & 25.53(4)  & -4.6(5)   \\
            &      &     &          &    &                \\
 $189\,{\rm GeV}$ & 29.73(14) &29.70(2)  & 27.98(2)  & 28.78(4)  & -3.2(5)     \\
            &      &     &      &    &                \\
 $200\,{\rm GeV}$ &36.45(23)  &35.93(4)  & 33.85(4)  & 34.97(6)  & -4.1(6)    \\
            &      &     &     &    &                \\
 \hline
 \end{tabular}
\caption{\label{tab6} Cross section [fb] for the process
$e^+e^- \to e^- \bar \nu_e \mu^+ \nu_{\mu}$.
$|\cos\theta_e| > 0.997$, 
$E_{\mu} > 15\,$GeV, $|\cos\theta_{\mu}| < 0.95$. No ISR.}
\end{center}
\end {table}

\clearpage 
A second problem, relevant when including QED radiation, is the choice of
the scales $q^2_i$ to be used in the Structure Function (SF)
formalism, schematically represented in equation (\ref{eq1})
\begin{equation}
d \sigma = \prod_i \int dx_i\,\, D({q^2_i},x_i)~d \sigma_0\,.
\label{eq1}
\end{equation}
The choice $q^2_i \sim s$ is proven to reproduce accurately 
the inclusive four-fermion cross sections, 
at least for $s$-channel dominated processes. 
  For $t$-channel dominated processes, such as single-$W$ production,
a different choice is more adequate, as studies of certain processes
have shown. When an exact first order QED radiative correction
calculation exists for a $t$-channel dominated process, then the
result can be compared to a Structure Function calculation
with a $q^2$ scale related to the virtuality of the exchanged
$t$-channel photon. With such a $q^2$ value the two kinds of
calculations agree for small angle Bhabha scattering \cite{sbhab}
and multi-peripheral two-photon processes \cite{isr2}, where the exact
calculation already exists for some time \cite{gge}. When no exact
first order QED correction calculation is available, the first order
soft correction may also serve as a guideline to determine $q^2$,
as proven by studies performed by {\tt GRACE} and {\tt SWAP}
 \cite{isr2,isr3}.
In {\tt NEXTCALIBUR}, the choice of the scale is performed automatically,
event by event, according to the selected final state, 
as shown in table \ref{tabnext}. The final state with 1 $e^-$ (or 1 $e^+$) 
is relevant for single-$W$ processes.
\begin{table}[thb]
\begin{center}
\begin{tabular}{|l||c|c|} \hline 
Final State          & $q^2_-$  & $q^2_+$ \\ \hline \hline
No $e^\pm$ & $s$     & $s$     \\ \hline              
1  $e^-$   & $|t_-|$ & $s$     \\ \hline               
1  $e^+$   & $s$     & $|t_+|$ \\ \hline               
1  $e^-$ and 1  $e^+$  & $|t_-|$    & $|t_+|$ \\ \hline               
2  $e^-$ and 2  $e^+$  & min($|t_-|$)    & min($|t_+|$) \\ \hline
\end{tabular}
\caption{\label{tabnext}The choice of the QED scale 
in {\tt NEXTCALIBUR}.
$q^2_\pm$ are the scales of the incoming $e^\pm$ while
$t_\pm$ represent the $t$-channel invariants obtained
by combining initial and final state $e^\pm$ momenta.
When two combinations are possible, as in the last entry of the table, 
that one with the minimum value of $|t|$ is chosen, event by event.}
\end{center}
\end{table}

A further problem is generating a $p_t$ spectrum for the 
photons. Usual solutions are the QED Parton Shower
approach (QEDPS) and  the use of $p_t$ dependent Structure Functions.

For example $p_t$ dependent SF are implemented in {\tt NEXTCALIBUR} via
the replacement
\begin{eqnarray}
\ln(\frac{q^2}{m^2_e})-1~~\to~~ 
\frac{1}{1-c+2\frac{m^2_e}{q^2}} -2 \frac{m^2_e}{q^2}
\frac{1}{(1-c+2\frac{m^2_e}{q^2})^2} \nonumber
\end{eqnarray}
in the  the soft part of the collinear SF, and
\begin{eqnarray}
\ln(\frac{q^2}{m^2_e})-1&\to& 
\frac{1}{1-c+2\frac{m^2_e}{q^2}} 
+\frac{1-x}{1+x} \cdot \frac{1}{2}
-4 \frac{m^2_e}{q^2}
\frac{1}{(1+x)(1-c+2\frac{m^2_e}{q^2})^2}
\nonumber
\end{eqnarray}
in the hard contributions.
After integrating over $c$, one gets $\ln(\frac{q^2}{m^2_e})-1$.
The inclusive QED result is therefore recovered and the photon 
spectrum is exact for small $p_t$.

Also {\tt GRACE} and {\tt SWAP} implement 
QEDPS and/or $p_t$ dependent SF.

\begin{figure}[thb]
\begin{center}
\SetWidth{1}
\begin{picture}(400,250)(-50,20)
\LinAxis(0,50)(300,50)(10,2,5,0,1.5)
\LinAxis(0,250)(300,250)(10,2,-5,0,1.5)
\LogAxis(0,50)(0,250)(4,-5,0,1.5)
\LogAxis(300,50)(300,250)(4,5,0,1.5)
\Text(0  ,40)[t]{$-1$}
\Text(150,40)[t]{$0$}
\Text(300,40)[t]{$1$}
\Text(-27,100)[l]{$10^{-1}$}
\Text(-27,150)[l]{$10^{0 }$}
\Text(-27,200)[l]{$10^{1 }$}
\Text(-60,227)[l]{$\frac{1}{\sigma} \frac{d\sigma}{d \cos \theta_\gamma}$}
\Text(150,20)[t]{$\cos \theta_\gamma$}
\Line(   0.0, 189.7)(   3.0, 189.7)
\Line(   3.0, 145.6)(   6.0, 145.6)
\Line(   6.0, 135.3)(   9.0, 135.3)
\Line(   9.0, 129.3)(  12.0, 129.3)
\Line(  12.0, 125.4)(  15.0, 125.4)
\Line(  15.0, 119.9)(  18.0, 119.9)
\Line(  18.0, 115.5)(  21.0, 115.5)
\Line(  21.0, 114.6)(  24.0, 114.6)
\Line(  24.0, 110.2)(  27.0, 110.2)
\Line(  27.0, 109.4)(  30.0, 109.4)
\Line(  30.0, 108.5)(  33.0, 108.5)
\Line(  33.0, 107.1)(  36.0, 107.1)
\Line(  36.0, 105.6)(  39.0, 105.6)
\Line(  39.0, 101.8)(  42.0, 101.8)
\Line(  42.0, 102.1)(  45.0, 102.1)
\Line(  45.0, 101.1)(  48.0, 101.1)
\Line(  48.0, 104.9)(  51.0, 104.9)
\Line(  51.0,  98.5)(  54.0,  98.5)
\Line(  54.0,  98.2)(  57.0,  98.2)
\Line(  57.0,  97.5)(  60.0,  97.5)
\Line(  60.0,  94.3)(  63.0,  94.3)
\Line(  63.0,  96.3)(  66.0,  96.3)
\Line(  66.0,  97.8)(  69.0,  97.8)
\Line(  69.0,  98.6)(  72.0,  98.6)
\Line(  72.0,  95.3)(  75.0,  95.3)
\Line(  75.0,  94.9)(  78.0,  94.9)
\Line(  78.0,  94.1)(  81.0,  94.1)
\Line(  81.0,  91.8)(  84.0,  91.8)
\Line(  84.0,  89.8)(  87.0,  89.8)
\Line(  87.0,  91.3)(  90.0,  91.3)
\Line(  90.0,  89.3)(  93.0,  89.3)
\Line(  93.0,  91.4)(  96.0,  91.4)
\Line(  96.0,  87.2)(  99.0,  87.2)
\Line(  99.0,  93.5)( 102.0,  93.5)
\Line( 102.0,  88.1)( 105.0,  88.1)
\Line( 105.0,  88.9)( 108.0,  88.9)
\Line( 108.0,  91.7)( 111.0,  91.7)
\Line( 111.0,  90.7)( 114.0,  90.7)
\Line( 114.0,  90.2)( 117.0,  90.2)
\Line( 117.0,  89.6)( 120.0,  89.6)
\Line( 120.0,  88.0)( 123.0,  88.0)
\Line( 123.0,  85.8)( 126.0,  85.8)
\Line( 126.0,  84.3)( 129.0,  84.3)
\Line( 129.0,  89.3)( 132.0,  89.3)
\Line( 132.0,  89.7)( 135.0,  89.7)
\Line( 135.0,  89.6)( 138.0,  89.6)
\Line( 138.0,  87.5)( 141.0,  87.5)
\Line( 141.0,  91.2)( 144.0,  91.2)
\Line( 144.0,  84.7)( 147.0,  84.7)
\Line( 147.0,  87.8)( 150.0,  87.8)
\Line( 150.0,  89.0)( 153.0,  89.0)
\Line( 153.0,  91.0)( 156.0,  91.0)
\Line( 156.0,  86.8)( 159.0,  86.8)
\Line( 159.0,  88.0)( 162.0,  88.0)
\Line( 162.0,  90.2)( 165.0,  90.2)
\Line( 165.0,  86.9)( 168.0,  86.9)
\Line( 168.0,  87.4)( 171.0,  87.4)
\Line( 171.0,  89.9)( 174.0,  89.9)
\Line( 174.0,  89.1)( 177.0,  89.1)
\Line( 177.0,  89.9)( 180.0,  89.9)
\Line( 180.0,  93.0)( 183.0,  93.0)
\Line( 183.0,  90.1)( 186.0,  90.1)
\Line( 186.0,  90.2)( 189.0,  90.2)
\Line( 189.0,  89.6)( 192.0,  89.6)
\Line( 192.0,  93.0)( 195.0,  93.0)
\Line( 195.0,  88.0)( 198.0,  88.0)
\Line( 198.0,  91.5)( 201.0,  91.5)
\Line( 201.0,  93.5)( 204.0,  93.5)
\Line( 204.0,  93.2)( 207.0,  93.2)
\Line( 207.0,  93.3)( 210.0,  93.3)
\Line( 210.0,  92.9)( 213.0,  92.9)
\Line( 213.0,  92.0)( 216.0,  92.0)
\Line( 216.0,  94.4)( 219.0,  94.4)
\Line( 219.0,  95.6)( 222.0,  95.6)
\Line( 222.0,  96.5)( 225.0,  96.5)
\Line( 225.0,  97.0)( 228.0,  97.0)
\Line( 228.0,  98.5)( 231.0,  98.5)
\Line( 231.0,  94.2)( 234.0,  94.2)
\Line( 234.0,  99.7)( 237.0,  99.7)
\Line( 237.0,  99.7)( 240.0,  99.7)
\Line( 240.0, 100.3)( 243.0, 100.3)
\Line( 243.0, 101.4)( 246.0, 101.4)
\Line( 246.0,  99.9)( 249.0,  99.9)
\Line( 249.0, 102.9)( 252.0, 102.9)
\Line( 252.0, 102.5)( 255.0, 102.5)
\Line( 255.0, 107.2)( 258.0, 107.2)
\Line( 258.0, 106.3)( 261.0, 106.3)
\Line( 261.0, 108.3)( 264.0, 108.3)
\Line( 264.0, 109.3)( 267.0, 109.3)
\Line( 267.0, 111.2)( 270.0, 111.2)
\Line( 270.0, 111.5)( 273.0, 111.5)
\Line( 273.0, 113.4)( 276.0, 113.4)
\Line( 276.0, 115.5)( 279.0, 115.5)
\Line( 279.0, 119.0)( 282.0, 119.0)
\Line( 282.0, 122.7)( 285.0, 122.7)
\Line( 285.0, 126.9)( 288.0, 126.9)
\Line( 288.0, 133.3)( 291.0, 133.3)
\Line( 291.0, 140.6)( 294.0, 140.6)
\Line( 294.0, 150.7)( 297.0, 150.7)
\Line( 297.0, 224.4)( 300.0, 224.4)
\Line(   3.0, 189.7)(   3.0, 145.6)
\Line(   6.0, 145.6)(   6.0, 135.3)
\Line(   9.0, 135.3)(   9.0, 129.3)
\Line(  12.0, 129.3)(  12.0, 125.4)
\Line(  15.0, 125.4)(  15.0, 119.9)
\Line(  18.0, 119.9)(  18.0, 115.5)
\Line(  21.0, 115.5)(  21.0, 114.6)
\Line(  24.0, 114.6)(  24.0, 110.2)
\Line(  27.0, 110.2)(  27.0, 109.4)
\Line(  30.0, 109.4)(  30.0, 108.5)
\Line(  33.0, 108.5)(  33.0, 107.1)
\Line(  36.0, 107.1)(  36.0, 105.6)
\Line(  39.0, 105.6)(  39.0, 101.8)
\Line(  42.0, 101.8)(  42.0, 102.1)
\Line(  45.0, 102.1)(  45.0, 101.1)
\Line(  48.0, 101.1)(  48.0, 104.9)
\Line(  51.0, 104.9)(  51.0,  98.5)
\Line(  54.0,  98.5)(  54.0,  98.2)
\Line(  57.0,  98.2)(  57.0,  97.5)
\Line(  60.0,  97.5)(  60.0,  94.3)
\Line(  63.0,  94.3)(  63.0,  96.3)
\Line(  66.0,  96.3)(  66.0,  97.8)
\Line(  69.0,  97.8)(  69.0,  98.6)
\Line(  72.0,  98.6)(  72.0,  95.3)
\Line(  75.0,  95.3)(  75.0,  94.9)
\Line(  78.0,  94.9)(  78.0,  94.1)
\Line(  81.0,  94.1)(  81.0,  91.8)
\Line(  84.0,  91.8)(  84.0,  89.8)
\Line(  87.0,  89.8)(  87.0,  91.3)
\Line(  90.0,  91.3)(  90.0,  89.3)
\Line(  93.0,  89.3)(  93.0,  91.4)
\Line(  96.0,  91.4)(  96.0,  87.2)
\Line(  99.0,  87.2)(  99.0,  93.5)
\Line( 102.0,  93.5)( 102.0,  88.1)
\Line( 105.0,  88.1)( 105.0,  88.9)
\Line( 108.0,  88.9)( 108.0,  91.7)
\Line( 111.0,  91.7)( 111.0,  90.7)
\Line( 114.0,  90.7)( 114.0,  90.2)
\Line( 117.0,  90.2)( 117.0,  89.6)
\Line( 120.0,  89.6)( 120.0,  88.0)
\Line( 123.0,  88.0)( 123.0,  85.8)
\Line( 126.0,  85.8)( 126.0,  84.3)
\Line( 129.0,  84.3)( 129.0,  89.3)
\Line( 132.0,  89.3)( 132.0,  89.7)
\Line( 135.0,  89.7)( 135.0,  89.6)
\Line( 138.0,  89.6)( 138.0,  87.5)
\Line( 141.0,  87.5)( 141.0,  91.2)
\Line( 144.0,  91.2)( 144.0,  84.7)
\Line( 147.0,  84.7)( 147.0,  87.8)
\Line( 150.0,  87.8)( 150.0,  89.0)
\Line( 153.0,  89.0)( 153.0,  91.0)
\Line( 156.0,  91.0)( 156.0,  86.8)
\Line( 159.0,  86.8)( 159.0,  88.0)
\Line( 162.0,  88.0)( 162.0,  90.2)
\Line( 165.0,  90.2)( 165.0,  86.9)
\Line( 168.0,  86.9)( 168.0,  87.4)
\Line( 171.0,  87.4)( 171.0,  89.9)
\Line( 174.0,  89.9)( 174.0,  89.1)
\Line( 177.0,  89.1)( 177.0,  89.9)
\Line( 180.0,  89.9)( 180.0,  93.0)
\Line( 183.0,  93.0)( 183.0,  90.1)
\Line( 186.0,  90.1)( 186.0,  90.2)
\Line( 189.0,  90.2)( 189.0,  89.6)
\Line( 192.0,  89.6)( 192.0,  93.0)
\Line( 195.0,  93.0)( 195.0,  88.0)
\Line( 198.0,  88.0)( 198.0,  91.5)
\Line( 201.0,  91.5)( 201.0,  93.5)
\Line( 204.0,  93.5)( 204.0,  93.2)
\Line( 207.0,  93.2)( 207.0,  93.3)
\Line( 210.0,  93.3)( 210.0,  92.9)
\Line( 213.0,  92.9)( 213.0,  92.0)
\Line( 216.0,  92.0)( 216.0,  94.4)
\Line( 219.0,  94.4)( 219.0,  95.6)
\Line( 222.0,  95.6)( 222.0,  96.5)
\Line( 225.0,  96.5)( 225.0,  97.0)
\Line( 228.0,  97.0)( 228.0,  98.5)
\Line( 231.0,  98.5)( 231.0,  94.2)
\Line( 234.0,  94.2)( 234.0,  99.7)
\Line( 237.0,  99.7)( 237.0,  99.7)
\Line( 240.0,  99.7)( 240.0, 100.3)
\Line( 243.0, 100.3)( 243.0, 101.4)
\Line( 246.0, 101.4)( 246.0,  99.9)
\Line( 249.0,  99.9)( 249.0, 102.9)
\Line( 252.0, 102.9)( 252.0, 102.5)
\Line( 255.0, 102.5)( 255.0, 107.2)
\Line( 258.0, 107.2)( 258.0, 106.3)
\Line( 261.0, 106.3)( 261.0, 108.3)
\Line( 264.0, 108.3)( 264.0, 109.3)
\Line( 267.0, 109.3)( 267.0, 111.2)
\Line( 270.0, 111.2)( 270.0, 111.5)
\Line( 273.0, 111.5)( 273.0, 113.4)
\Line( 276.0, 113.4)( 276.0, 115.5)
\Line( 279.0, 115.5)( 279.0, 119.0)
\Line( 282.0, 119.0)( 282.0, 122.7)
\Line( 285.0, 122.7)( 285.0, 126.9)
\Line( 288.0, 126.9)( 288.0, 133.3)
\Line( 291.0, 133.3)( 291.0, 140.6)
\Line( 294.0, 140.6)( 294.0, 150.7)
\Line( 297.0, 150.7)( 297.0, 224.4)
\end{picture}
\caption{\label{distnext} $\cos \theta_\gamma$ distribution for
$~e^+ e^- \to e^- \bar \nu_e u \bar d\, (\gamma)$, 
by {\tt NEXTCALIBUR}.}
\end{center}
\end{figure}
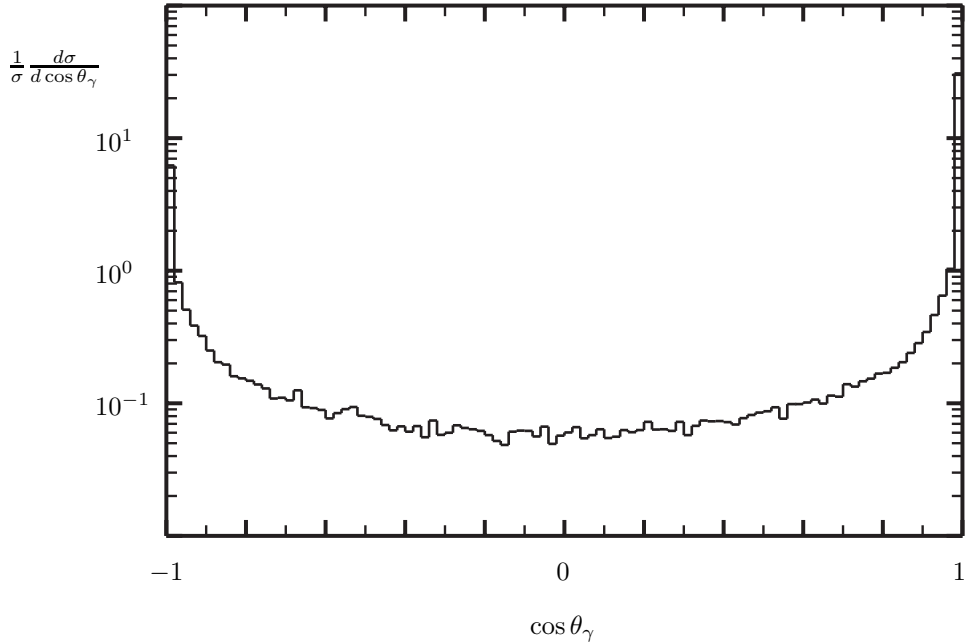

Since $q^2=s$ for $s$-channel legs and 
$q^2=|t|$ for $t$-channel dominated legs,
the $p_t$ distribution is different in the two cases.
In order to show this effect, the distribution 
in $\cos \theta_\gamma$ for the most energetic photon
(with respect to the incoming $e^+$) in
$~e^+ e^- \to e^- \bar \nu_e u \bar d\, (\gamma)$, is shown
in figure \ref{distnext} as predicted by {\tt NEXTCALIBUR}.

The shape of the distribution can be easily understood 
in terms of radiation suppression.
In fact $|t| \to 0$ implies $D(|t|,x)~\sim \delta(1-x)$.
Therefore the emission from the incoming $e^-$
is suppressed, and the photons are preferably emitted 
collinearly by the incoming $e^+$.

The question of comparing QEDPS and SF has been
addressed by the {\tt GRACE} group. The differences between the
two approaches are at the order of 1\% for single-$W$
processes. Furthermore, the $E_\gamma$ distribution is well 
reproduced for soft photons, with respect to an exact 
Matrix Element calculations, while 20\% differences are
observed (and expected) in the hard region. 

In conclusion, putting together all the effects, 
a theoretical error of 5\% \footnote{For semi-leptonic
processes this number refers to the High Mass region, 
namely to situations when appropriate cuts on 
the jet-jet system are applied in order to
suppress the non-perturbative resolved photon contributions.}
is assigned to single-$W$ processes. 
This result has to be compared with the accuracy required by
the LEP2 collaborations namely {2\%} for $e^+ e^-\to e^-~\bar \nu_e ~q~\bar q^\prime~(\gamma)$,~{5\%} for $e^+ e^-\to e^-~\bar \nu_e ~e^+  ~\nu_e    ~(\gamma)$,~{5\%} for 
$e^+ e^-\to e^-~\bar \nu_e ~\mu^+~\nu_{\mu}~(\gamma)$ \cite{wshop}.
However, different pieces of knowledge are 
still scattered in different codes, and an improvement
of the present situation is possible via a multi-step 
experimental procedure, in which, for example, 
code A is used to correct code B for the missing effects. 

At the LC further improvements are needed. 
In particular, the genuine EW corrections get larger with 
increasing energy.
Again, a full one-loop four-fermion calculation seems unavoidable,
for high precision measurements.
\section{Four fermions plus 1 visible photon} 
This signature gives information on the quartic gauge coupling
(see figure \ref{quartic}) and is relevant when studying 
processes with three final state bosons, such as
$W^+ W^- \gamma$ production, $Z Z \gamma$ and $Z \gamma \gamma$.
Furthermore, it is a building block for the full computation of
$e^+ e^- \to$ 4f at ${\cal O}({\alpha})$.

A bunch of codes contributed, with different strategies.
{\tt CompHEP}, {\tt GRACE} and {\tt HELAC/PHEGAS} compute the 
exact Matrix Element (ME) with massive fermions.
{\tt RACOONWW} uses exact ME, but in the limit of massless fermions.
{\tt NEXTCALIBUR} generates photons only through $p_t$ dependent SF.
{\tt WRAP} has a matching between ME and SF generated photons. 
The last approach allows to estimate the size of the double 
counting when blindly dressing the $4f + \gamma$ ME 
with collinear ISR.  {\tt WRAP} observed effects up to 5\%, depending
on the energy cut used to define the visible photon.

In figure \ref{chep} a study by
{\tt CompHEP} is shown on the reliability of 
a narrow-width approximation.

Distributions generated by {\tt GRACE} are shown in figure \ref{grcfig}.

\begin{figure}[thb]
\begin{center}
\begin{picture}(200,80)
\SetScale{0.7}
\SetWidth{1}
\sof(50,45)
\flin{0,15}{15,-5} \flin{15,-5}{0,-25}
\glin{15,-5}{30,-5}{3}
\glin{40,30}{30,-5}{4}
\glin{30,-5}{40,-40}{4}
\glin{30,-5}{70,-5}{5}
\Text(57,-7.5)[l]{{$\gamma$}}
\Text(14,10)[b]{{$W^-$}}
\Text(14,-20)[t]{{$W^+$}}
\flin{55,10}{40,30} \flin{40,30}{55,50}
\flin{40,-40}{55,-20} \flin{55,-60}{40,-40}
\end{picture}
\caption{\label{quartic} Quartic gauge boson coupling in four-fermion
production plus 1 visible $\gamma$.}
\end{center}
\end{figure}
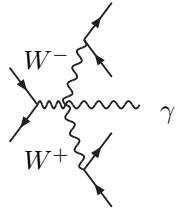
\begin{figure}[thb]
\vskip 0.5cm
\psfig{figure=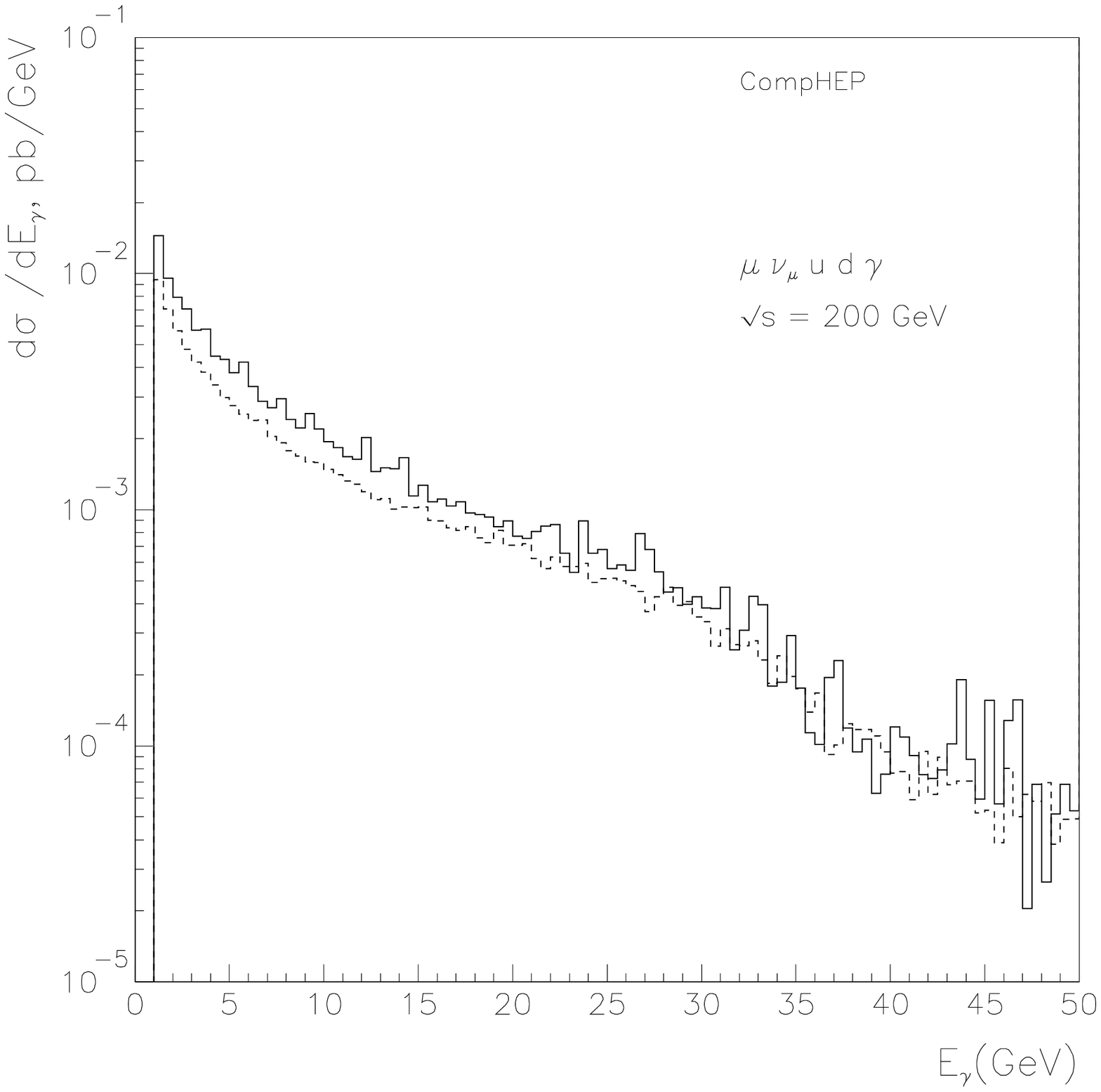,height=5.cm}
\vskip -5.cm
\hskip 9cm
\psfig{figure=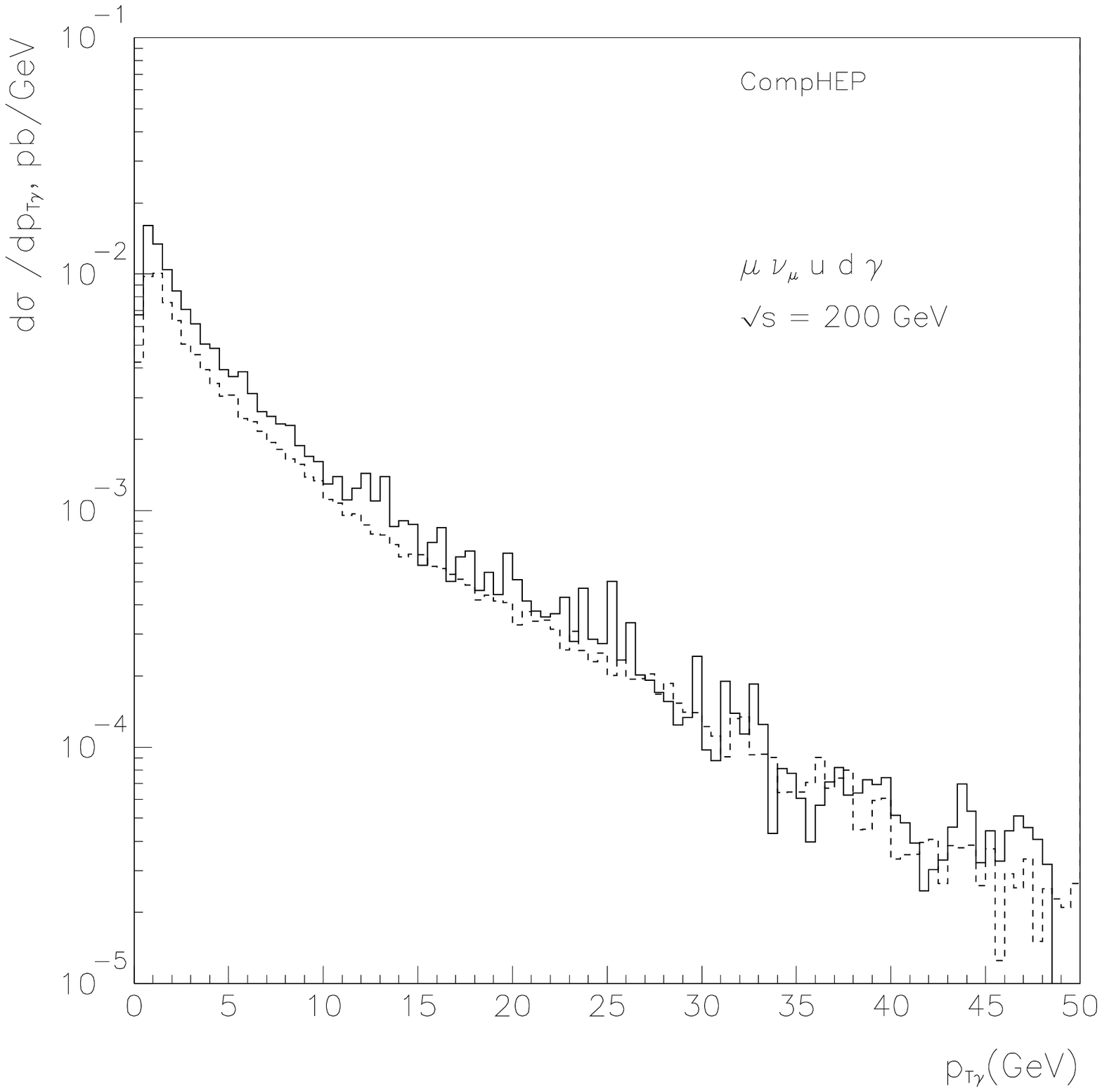,height=5.cm}
\psfig{figure=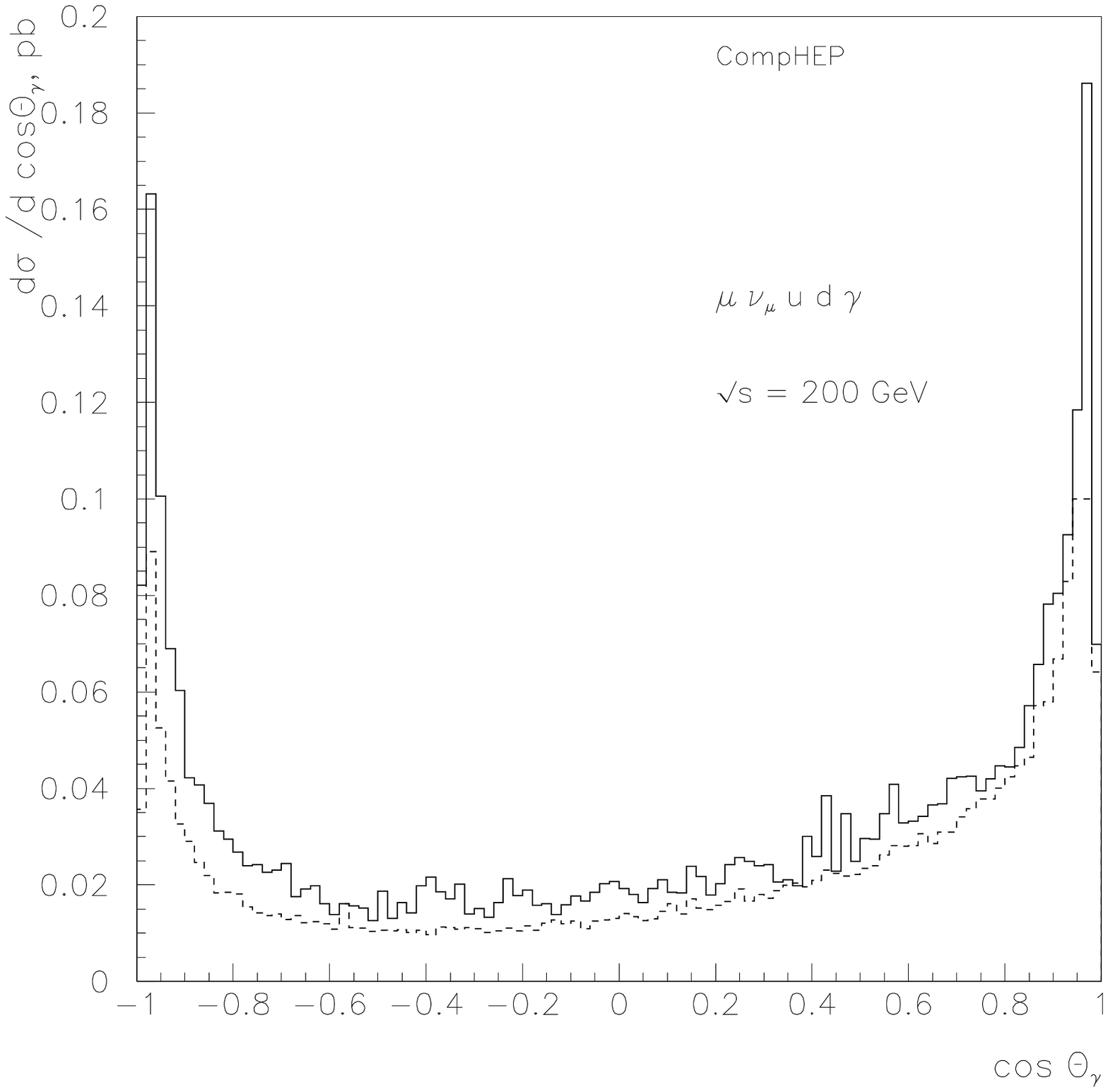,height=5.cm}
\vskip -5.cm
\hskip 9cm
\psfig{figure=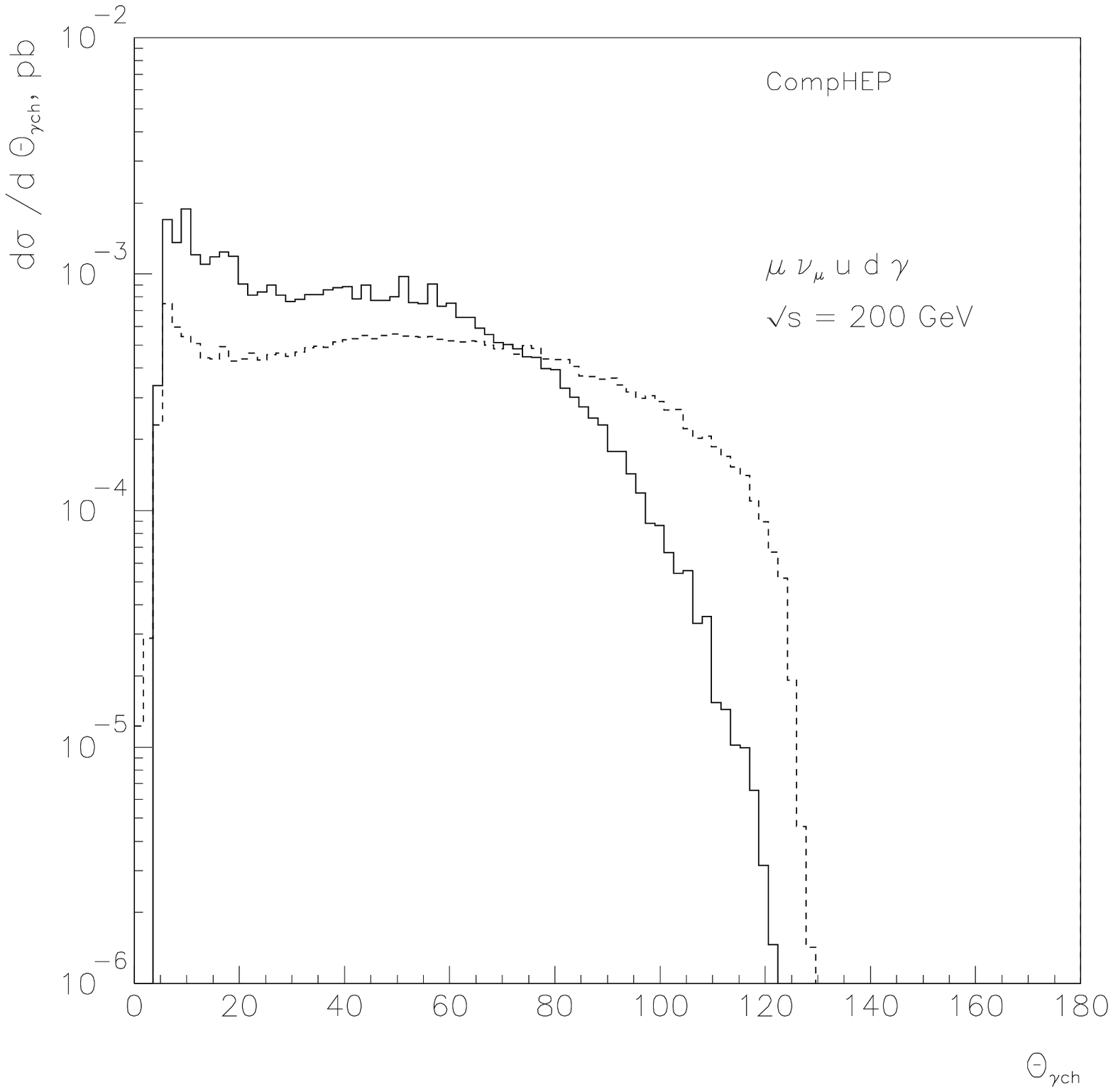,height=5.cm}
\caption{\label{chep} Distributions in the $\gamma$ energy, $\gamma$ transverse momentum,
$\gamma$ angle with the beam, and in the opening angle between the
$\gamma$ and the nearest charged fermion. 
The distributions for the
$e^+ e^- \to \gamma \mu \bar \nu_\mu u \bar d$ are shown by the solid
line and the distributions for the $e^+ e^- \to \gamma \mu \bar
\nu_\mu W^+$ with the following $W$ isotropic decay are shown by
the dashed line.}
\end{figure}
\clearpage

\begin{figure}[thb]
\psfig{figure=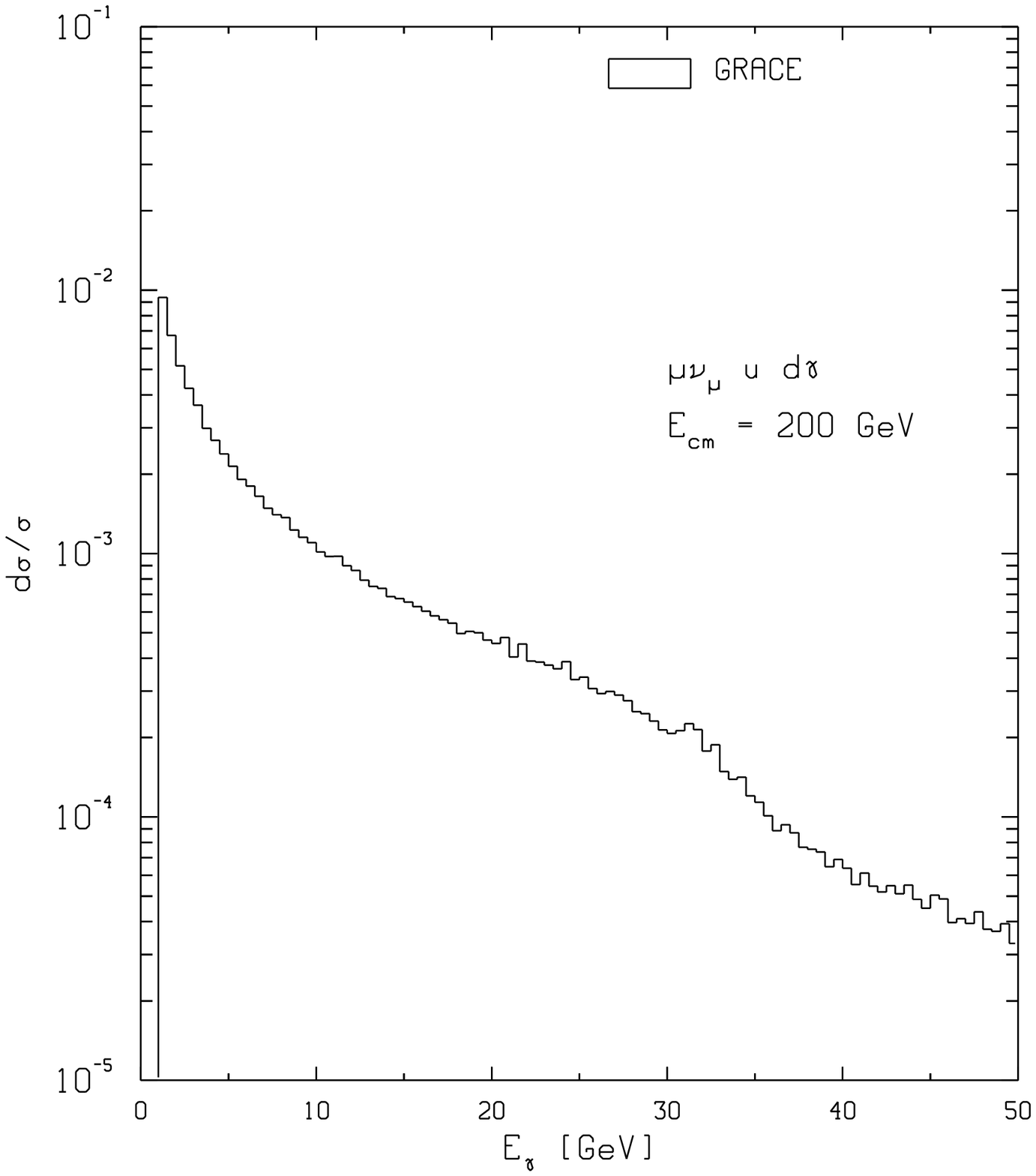,height= 9cm}
\vskip -9.cm
\hskip 9cm
\psfig{figure=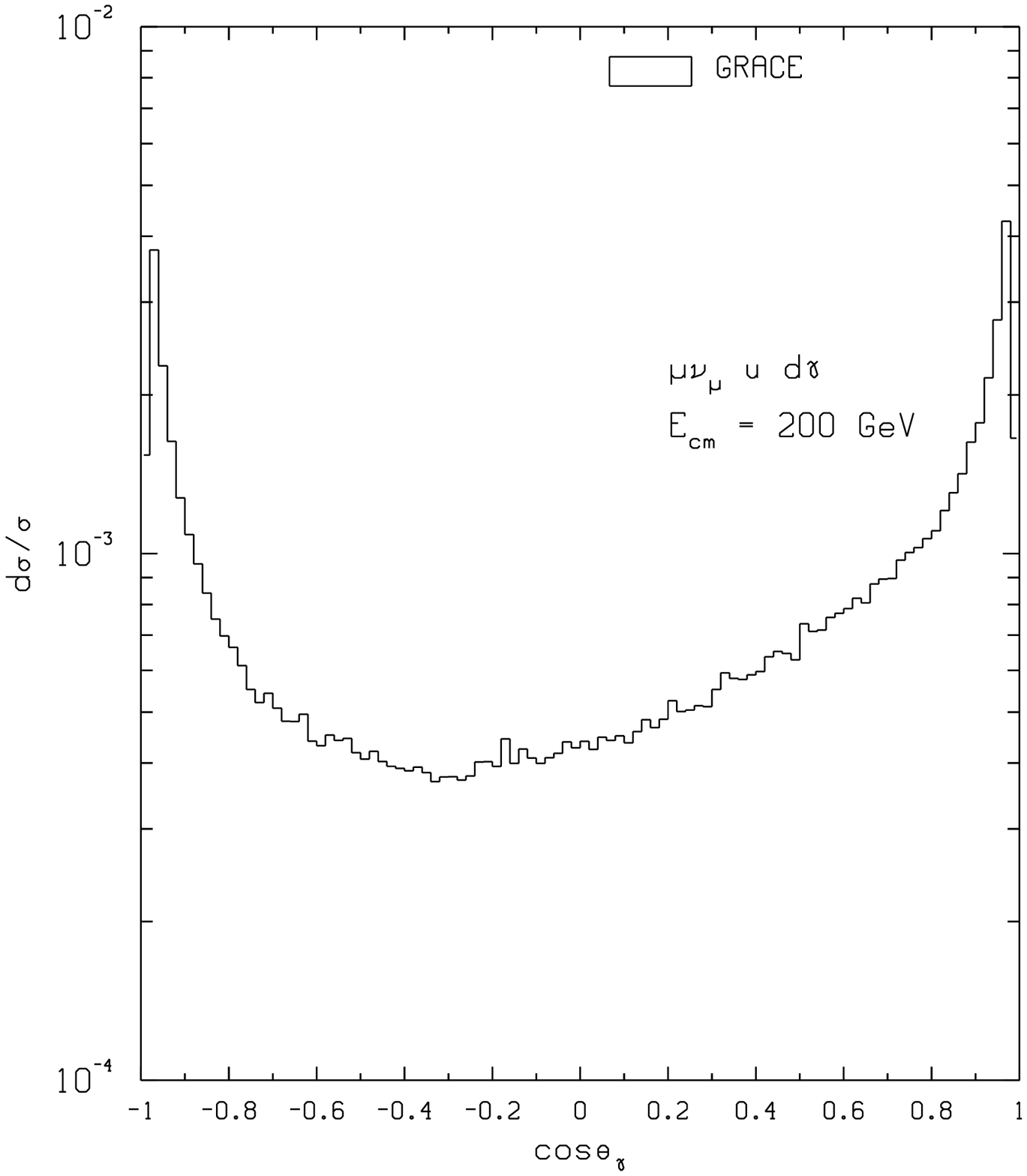,height= 9cm}
\vskip -1.cm
\psfig{figure=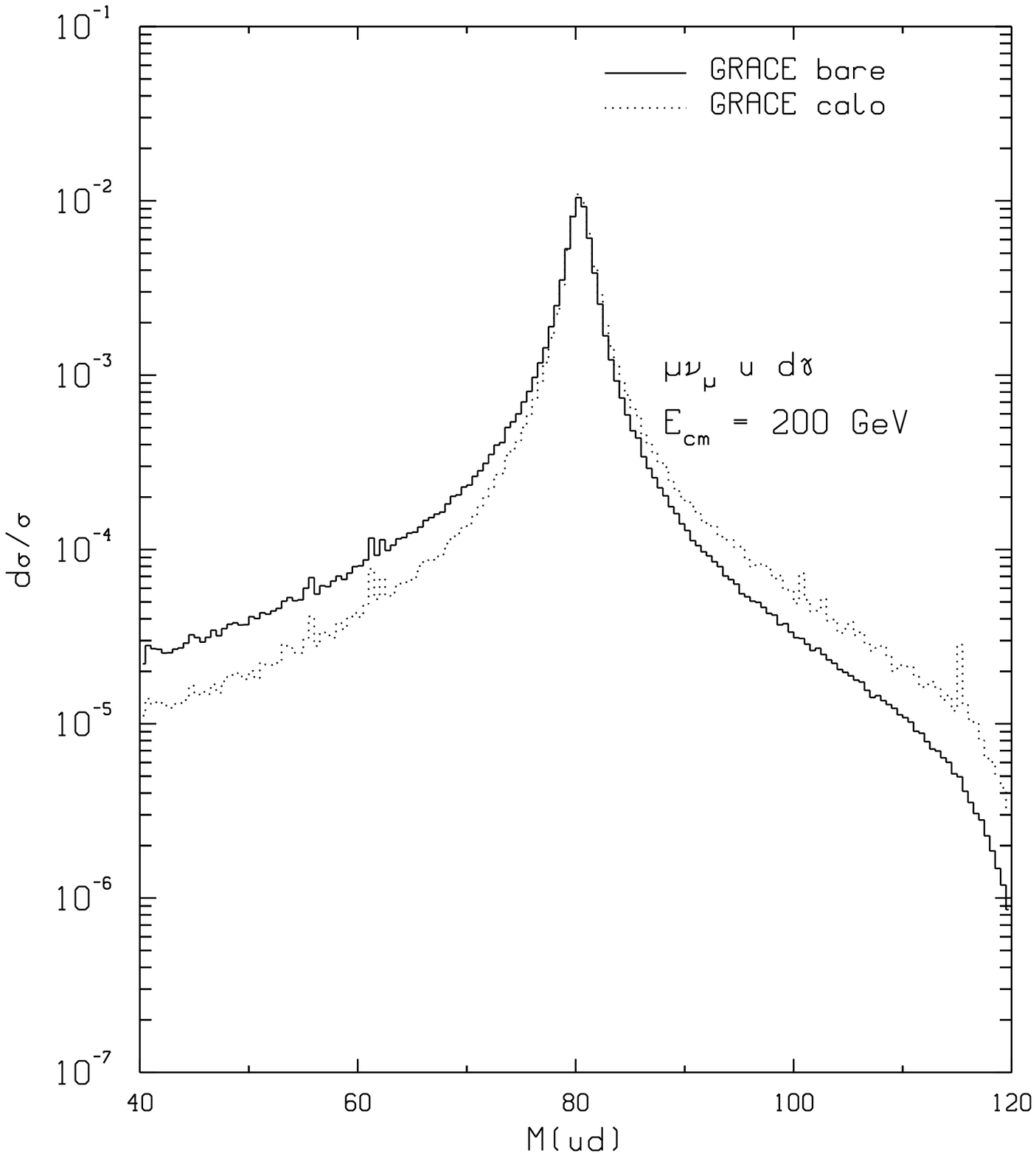,height= 9cm}
\vskip -9.cm
\hskip 9cm
\psfig{figure=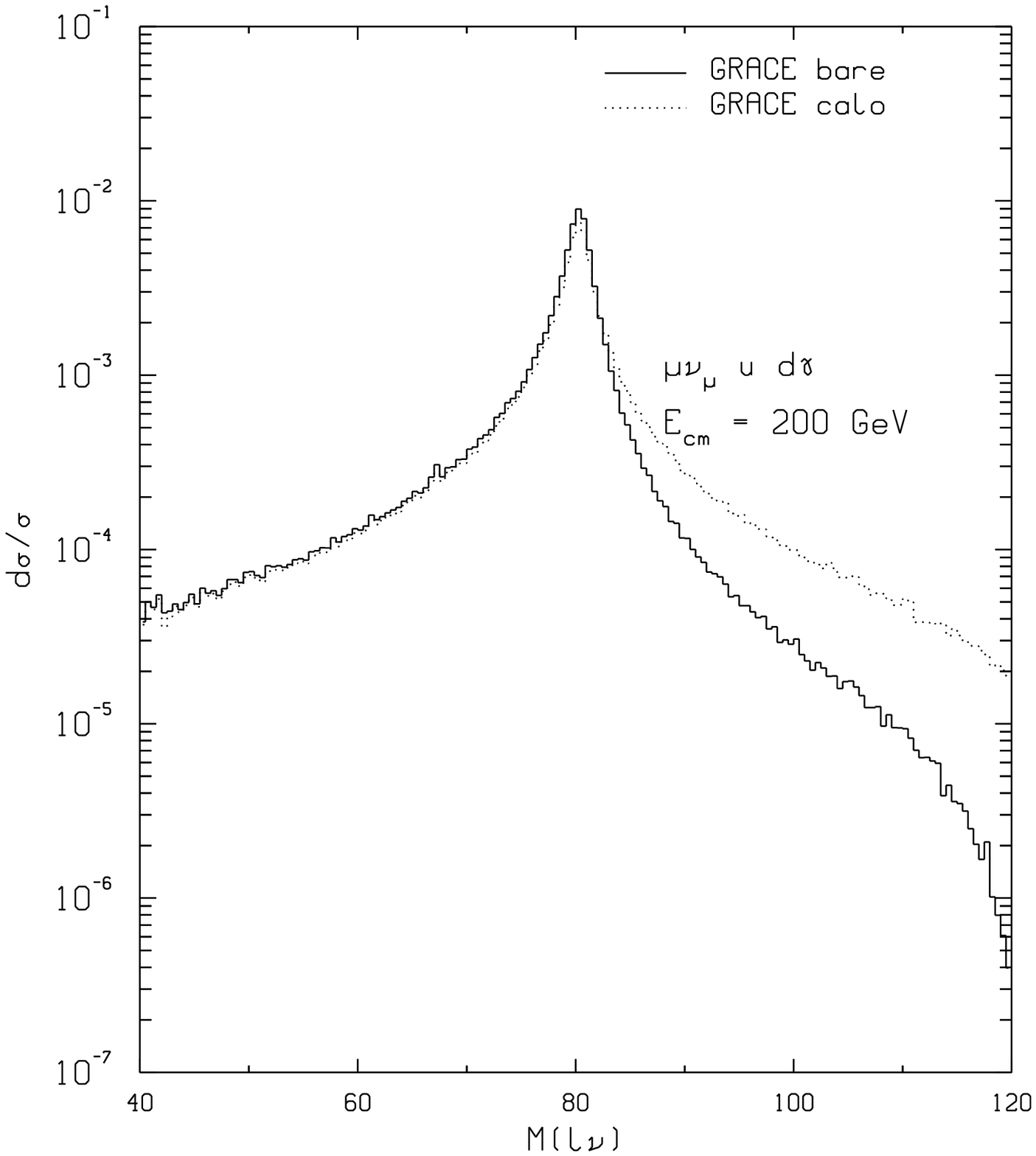,height= 9cm}
\caption{\label{grcfig} $E_{\gamma}$, $\cos \theta_{\gamma}$ and invariant
mass distributions for $\mu \nu_\mu u d \gamma$.}
\end{figure}
\clearpage

In tables \ref{nextab1} and \ref{nextab2} the cross sections for
inclusive ($\sigma_{\rm tot}$), non radiative ($\sigma_{\rm nrad}$) , 
single radiative ($\sigma_{\rm srad}$) 
and double radiative ($\sigma_{\rm drad}$) events 
are shown, as computed by {\tt NEXTCALIBUR}.
This kind of study is useful for signal definition.
\begin{table}[thb]
\begin{center}
\begin{tabular}{|c|c|}
\hline
&\\
Type & Cross-section \\
&\\
\hline
&\\
$\sigma_{\rm tot}$    &   16.107(9)    \\
$\sigma_{\rm nrad}$   &   15.018(9)    \\
$\sigma_{\rm srad}$   &    1.0697(30)  \\
$\sigma_{\rm drad}$   &    0.0189(4)   \\
\hline
\end{tabular}
\vskip 0.4cm
\caption{\label{nextab1}$\sigma $ in fb for $e^+(1) e^-(2) 
\to \mu^-(3) \mu^+(4) u(5) \bar u(6)$.
$M(34) > 10\,$GeV and $M(56) > 10\,$GeV.
$ZZ$ like cuts, $E_{\gamma} > 1\,{\rm GeV}, 
|\cos\theta_{\gamma}| < 0.985$.}
\end{center}
\end{table}
\begin{table}[thb]
\begin{center}
\begin{tabular}{|c|c|}
\hline
&\\
Type & Cross-section \\
&\\
\hline
&\\
$\sigma_{\rm tot}$    &   617.27(59) \\
$\sigma_{\rm nrad}$   &   578.19(58) \\
$\sigma_{\rm srad}$   &   38.54(16)  \\
$\sigma_{\rm drad}$   &    0.54(2)   \\
\hline
\end{tabular}
\vskip 0.4cm
\caption{\label{nextab2} $\sigma$ in fb for $e^+(1) e^-(2) 
\to \mu^-(3) \bar \nu_{\mu}(4) u(5) \bar d(6)$.
$M(56) > 10\,$GeV. $WW$ like cuts, $E_{\gamma} > 1\,{\rm GeV}, 
|\cos\theta_{\gamma}| < 0.985$.} 
\end{center}
\end{table}

\begin{figure}[thb]
\begin{center}
\psfig{figure=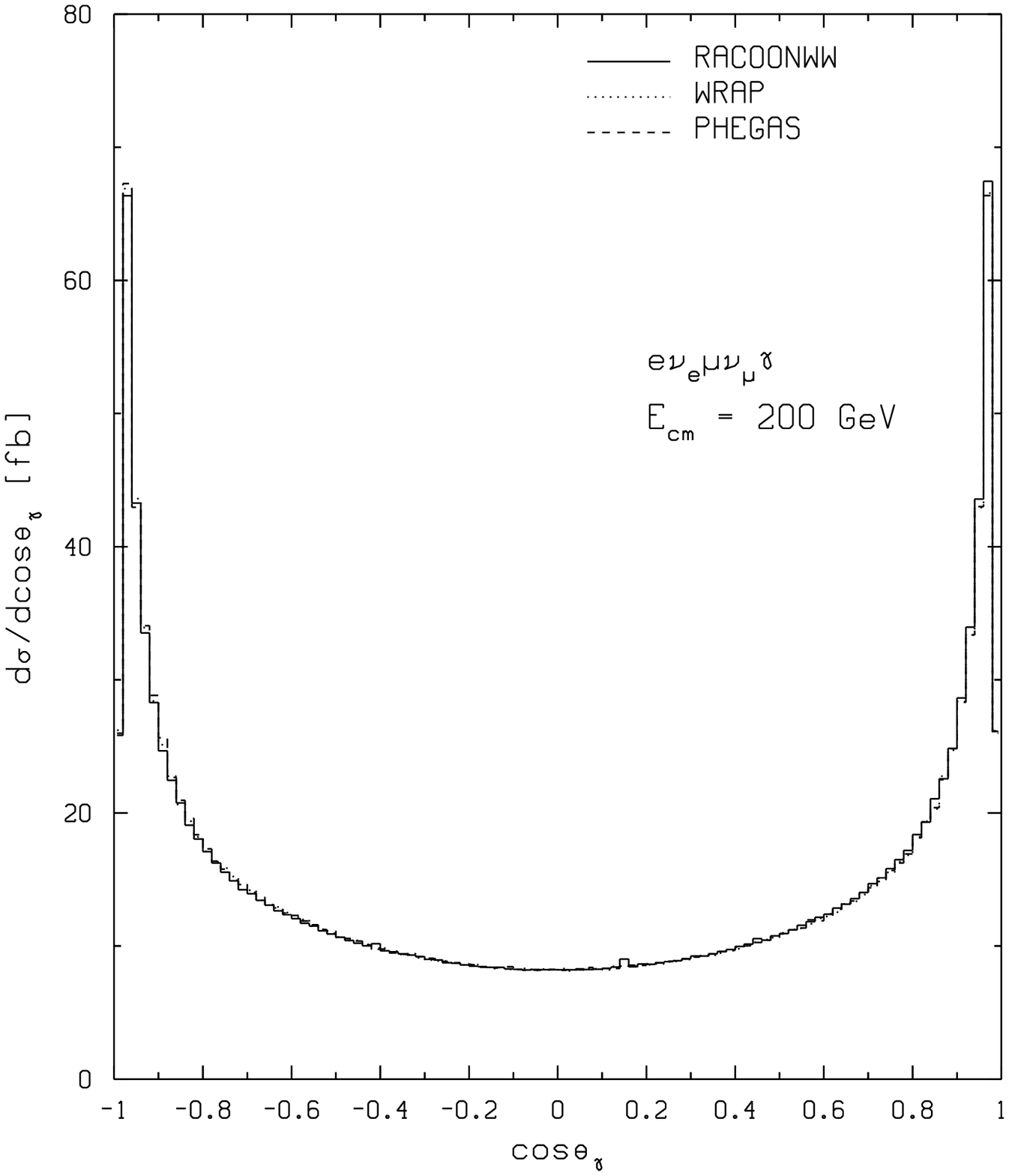,height= 8cm}
\vskip -8cm
\hskip  5cm
\psfig{figure= 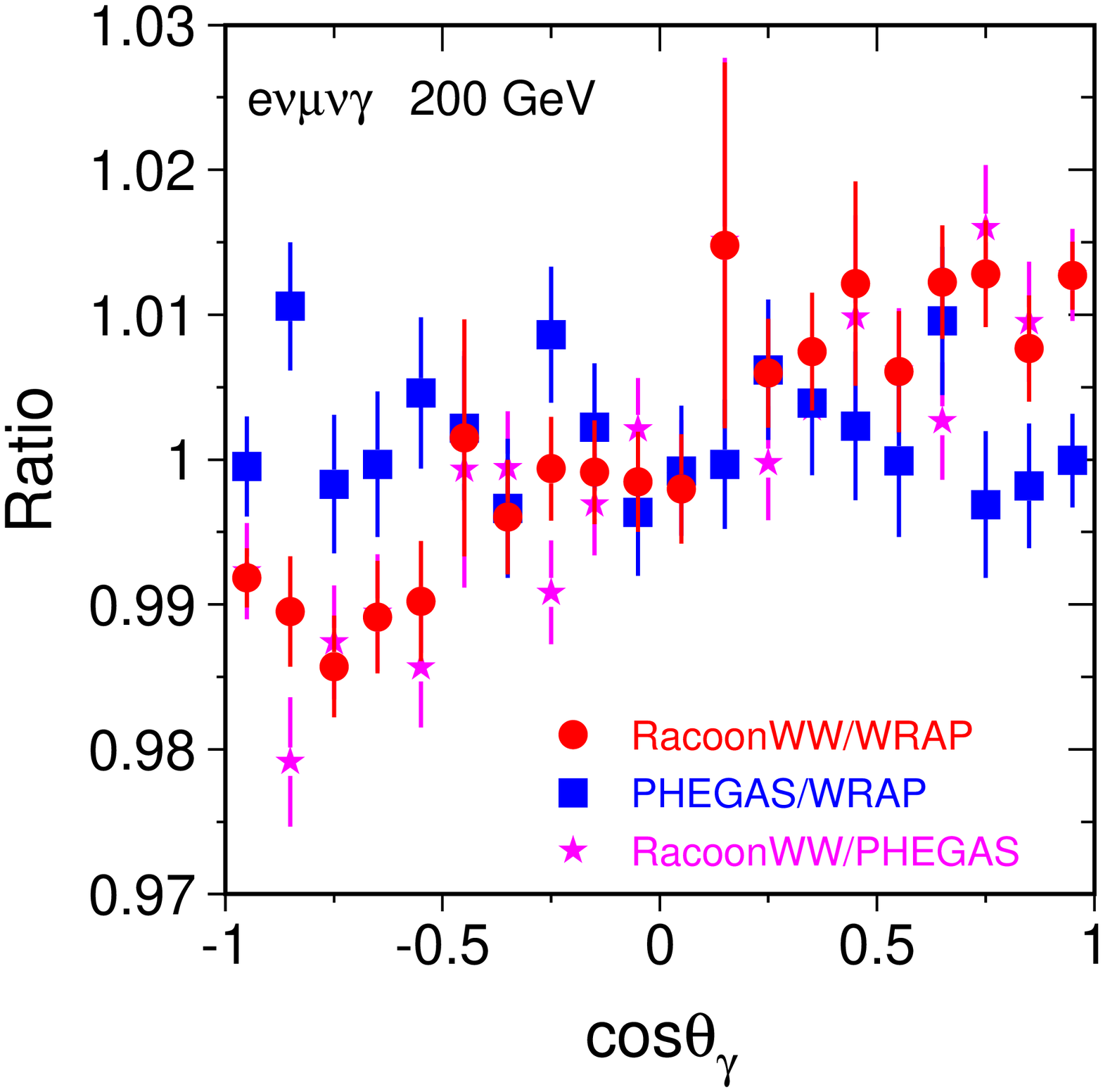,height= 8cm}
\vskip -0.5cm
\caption{\label{comp2} 
$\cos \theta_{\gamma}$ distributions and ratios for
$\nu_\mu \mu^+ e^- \bar \nu_e\gamma$.}
\end{center}
\end{figure}

 Finally, in table \ref{comp1} and figure \ref{comp2}, 
I show results of tuned comparisons among {\tt HELAC/PHEGAS}, 
{\tt RACOONWW} and {\tt WRAP}.
\begin{table}[thb]
\begin{center}
\begin{tabular}{|c|c|c|c|}
\hline
Process & {\tt WRAP} & {\tt RacoonWW} & {\tt PHEGAS/HELAC} \\
\hline
&&&\\
$u \bar d \mu^- \bar \nu_{\mu}\gamma$      
& 75.732(22) & 75.647(44) & 75.683(66)\\
$u \bar d e^- \bar \nu_e\gamma$             
& 78.249(43) & 78.224(47) & 78.186(76)\\
$\nu_\mu \mu^+ \tau^- \bar \nu_{\tau}\gamma$ 
& 28.263(9)  & 28.266(17) & 28.296(22) \\
$\nu_\mu \mu^+ e^- \bar \nu_e\gamma$     
& 29.304(19) & 29.276(17) & 29.309(25)\\
$u \bar d s \bar c\gamma$                  
&199.63(10)  &199.60(11) & 199.75(16) \\
&&&\\
\hline
\end{tabular}
\caption{\label{comp1}
$\sigma$ in fb from {\tt WRAP}, {\tt RacoonWW} 
and {\tt PHEGAS/HELAC}.}
\end{center}
\end{table}

\begin{table}[thb]
\begin{center}
\begin{tabular}{|c|c|c|c|}
\hline
channel & {\tt YFSZZ} & {\tt ZZTO} $G_F$-scheme & {\tt ZZTO} $\alpha$-scheme \\
\hline
\hline
$qqqq$            &    294.6794(490) &  298.4411(60)  &    294.5715(59)   \\
$qq\nu\nu$        &    175.4404(302) &  175.5622(35)  &    174.9855(35)   \\
$qq{\rm ll}$      &     88.1805(134) &   88.7146(18)  &     87.9881(18)   \\
${\rm ll}\nu\nu$  &     26.2530(463) &   26.0940(5)   &     26.1342(5)    \\
${\rm llll}$      &      6.5983(15)  &    6.5929(1)   &      6.5706(1)    \\
$\nu\nu\nu\nu$    &     26.1080(71)  &   25.8192(5)   &     25.9868(5)
\\
\hline
total             &    617.2596(755) &  621.2241(124) &    616.2366(123)  \\
\hline
\end{tabular}
\vskip 0.3cm
\caption{\label{tabzz} 
$ZZ$ cross-section [fb] at $\sqrt{s}= 188.6\,$GeV.
YFS exponentiation is used by {\tt YFSZZ}, while {\tt ZZTO} 
includes EFL corrections in the $\alpha$-scheme.} 
\end{center}
\end{table}
In conclusion, a very good technical precision has been reached
in the computation of four-fermion processes plus 1
additional photon. However, the 
non-logarithmic ${\cal O}(\alpha)$ corrections
are not known. Therefore a 2.5\% theoretical accuracy 
on total cross section and inclusive distributions is estimated
at LEP2 energies \cite{wshop}. Larger effects are expected at the LC.
\section{$Z$-pair production}
Less accuracy is required at LEP2 for this observable with respect 
to the $W$-pair case.
 Thought feasible in principle, a DPA $ZZ$ calculation 
is not available yet. A theoretical accuracy of 2\% on
$\sigma_{ZZ}$ is estimated at LEP2 by varying
the renormalization scheme and by comparing different codes 
and different treatments of the QED radiation (see table \ref{tabzz}).

\clearpage
\section{Conclusions}
Four-fermion Physics at LEP2 is in a good shape. 
Thanks to the results of ref. \cite{wshop}, improved
calculations are available for
\begin{itemize}
\item $W$-pair production 
\item Single-$W$ production
\item Four fermions plus 1 visible $\gamma$
\item $Z$-pair production.
\end{itemize}
In general, the theoretical accuracies required by the LEP2 experiments
are achieved.

At the LC more accuracy is needed. In particular 
radiative corrections must be included.

Four-fermion loop calculations at ${\cal O}(\alpha_{s})$ already
exist \cite{loopqcd}, while a full ${\cal O}(\alpha)$ EW
calculation is still missing.

\end{document}